# Nanoscale and Element-Specific Lattice Temperature Measurements using Core-Loss Electron Energy-Loss Spectroscopy

Levi D. Palmer,[1] Wonseok Lee,[1] Daniel B. Durham,[2] Javier Fajardo, Jr.,[3] Yuzi Liu,[2] A. Alec Talin,[3] Thomas E. Gage,[2,*] Scott K. Cushing[1,*]

[1]*Division of Chemistry and Chemical Engineering, California Institute of Technology, Pasadena, CA 91125, USA*
[2]*Center for Nanoscale Materials, Argonne National Laboratory, Lemont, IL 60439, USA*
[3]*Sandia National Laboratories, Livermore, California 94551, USA*

*Correspondence to Thomas E. Gage & Scott K. Cushing

## ABSTRACT

Measuring nanoscale local temperatures, particularly in vertically integrated and multi-component systems, remains challenging. Spectroscopic techniques like X-ray absorption and core-loss electron energy-loss spectroscopy (EELS) are sensitive to lattice temperature, but understanding thermal effects is nontrivial. This work explores the potential for nanoscale and element-specific core-loss thermometry by comparing the Si $L_{2,3}$ edge's temperature-dependent redshift against plasmon energy expansion thermometry (PEET) in a scanning TEM. Using density functional theory (DFT), time-dependent DFT, and the Bethe-Salpeter equation, we *ab initio* model both the Si $L_{2,3}$ and plasmon redshift. We find that the core-loss redshift occurs due to bandgap reduction from electron-phonon renormalization. Our results indicate that despite lower core-loss signal intensity compared to plasmon features, core-loss thermometry has key advantages and can be more accurate through standard spectral denoising. Specifically, we show that the Varshni equation easily interprets the core-loss redshift for semiconductors, which avoids plasmon spectral convolution for PEET in complex junctions and interfaces. We also find that core-loss thermometry is more accurate than PEET at modeling thermal lattice expansion in semiconductors, unless the specimen's temperature-dependent dielectric properties are fully characterized. Furthermore, core-loss thermometry has the potential to measure nanoscale heating in multi-component materials and stacked interfaces with elemental specificity at length scales smaller than the plasmon's wavefunction.

## INTRODUCTION

Accurate temperature measurements at the nanoscale are essential for understanding and managing thermal gradients in modern electronic, photonic, and energy conversion/storage devices. During operation, local thermal variations significantly influence device performance and reliability. For example, the breakdown of Dennard scaling for transistors and the increasing challenge of heat removal have limited transistor density and clock speeds in central processing units since the early 2000s.[1] Similarly, semiconductor devices used in transportation, renewable energy systems, and the electric grid require precise thermal control to ensure efficient switching and prevent degradation.[2] In contrast, modest temperature increases in semiconductor-based photocatalytic systems have recently been reported to substantially improve solar hydrogen generation.[3,4] Understanding when local heating is detrimental or beneficial to each application requires robust and accurate *in situ* temperature measurements with nanometer spatial resolution.

Plasmon energy expansion thermometry (PEET) is one technique that, in principle, enables sub-nm thermal imaging.[5–10] PEET measures temperature-induced shifts in a material's volume plasmon energy according to the free electron model following thermal expansion. In a scanning TEM (STEM) with electron energy-loss spectroscopy (EELS), PEET has been used to image nanoscale temperatures and thermal lattice expansion coefficients for nanoparticles, membranes, and two-dimensional materials.[5–11] Time-resolved PEET is also possible in optically-coupled electron microscopes.[12] However, recent studies show that the volume plasmon can vary with sample thickness and local dielectric properties.[13] The resolution in momentum ($k$) space can also obscure accurate PEET measurements because plasmon energies typically vary quadratically with momentum.[14–16] Local strain can also exceed the thermal expansion, especially at junctions and interfaces.[17] PEET also has difficulty measuring temperature in materials with small thermal expansion coefficients that result in <0.1 meV/K plasmon energy shifts. PEET's challenges necessitate additional spectromicroscopy tools for nanoscale thermal imaging.[18]

We propose core-loss thermometry as a complementary or alternative method to PEET that also has expanded utility at X-ray beamlines. Core-loss edges, such as those measured by core-loss EELS and X-ray absorption, are sensitive to changes in a material's electronic structure and valence states. While extended X-ray absorption fine structure (EXAFS) spectra can quantify lattice temperatures, its need for high-energy core-level transitions prohibit the signal and spatial resolution for core-loss EELS.[19,20] Ultrafast and photoexcited X-ray spectroscopy studies have found that core-level edges of semiconductors redshift with temperature, but the exact redshift mechanism is not well understood.[21–23] We aim to understand and predict this temperature-dependent redshift using EELS simulations to develop core-loss thermometry.

In this work, we use *in situ* STEM-EELS with controlled temperature-dependent low- and core-loss EEL spectra across multiple thickness regions of a crystalline Si lamella with native oxide. We evaluate PEET's and core-loss thermometry's accuracy across different thickness regions of the lamella and extract temperature-induced shifts of the Si $L_{2,3}$ edge ($2p \rightarrow 3s+d$ transition). To interpret these shifts, we incorporate *ab initio* simulations of the Si plasmon and $L_{2,3}$ edge using density functional theory (DFT), linearized time-dependent DFT (TDDFT), and the Bethe-Salpeter equation (BSE). Our analysis finds that the core-loss spectrum redshifts according to the reduction of Si temperature-dependent bandgap as predicted by the Varshni equation.[24,25] Compared to PEET, core-loss thermometry of semiconductors is less sensitive to effects from multiple plasmons, multiple scattering, temperature-dependent strain, small thermal expansion coefficients, surface contamination, and multi-element stacked junctions. We also find core-loss thermometry more accurate than PEET at measuring the thermal expansion coefficient when the temperature-dependent dielectric constant and environment are not fully characterized. Further, a standard spectral denoising algorithm can improve core-loss thermometry's accuracy. Ultimately, core-loss thermometry is a promising route for element-specific, nanoscale temperature mapping during operando or ultrafast measurements.

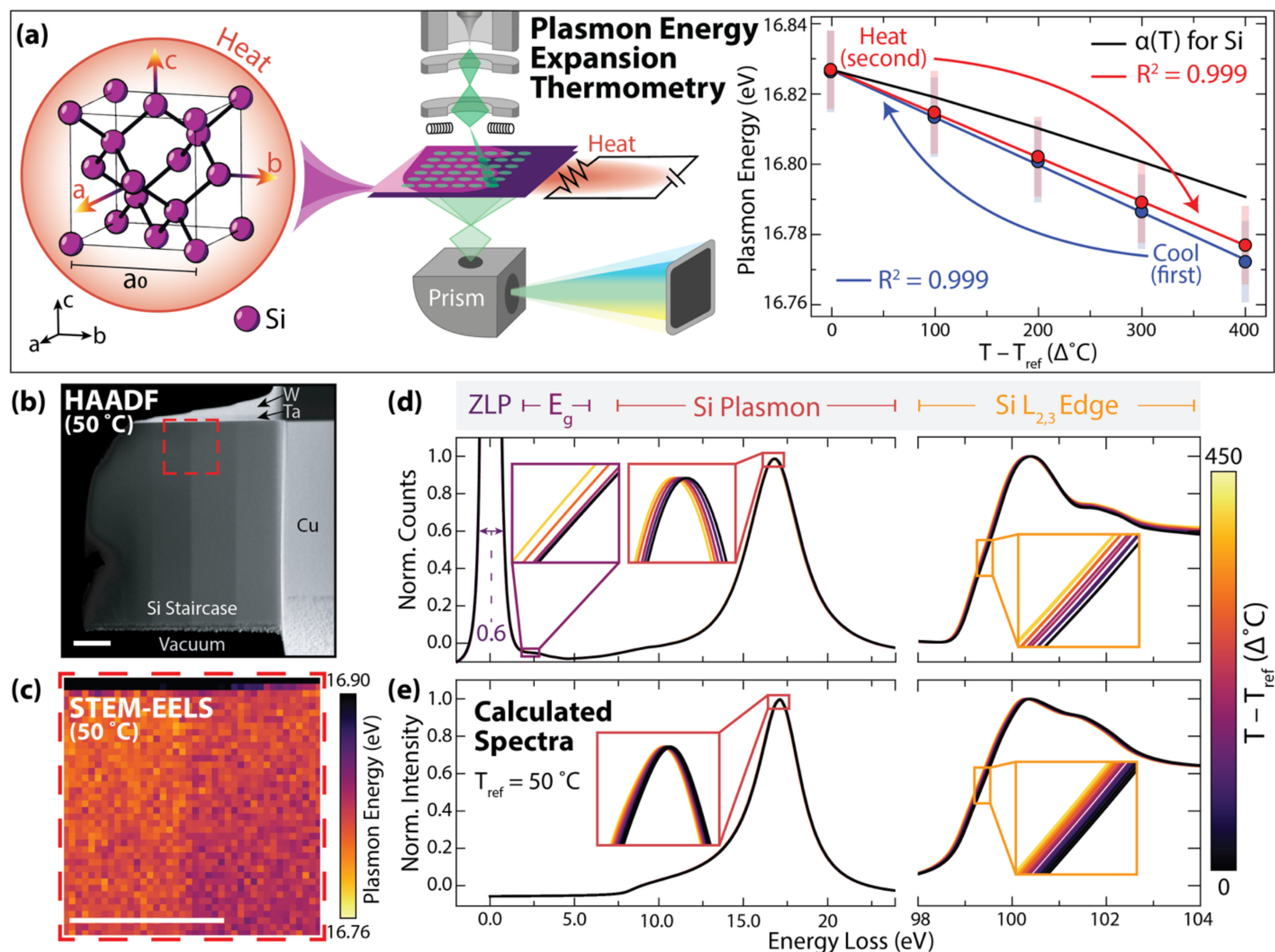

**Figure 1. Temperature-dependent scanning TEM electron energy-loss spectroscopy (STEM-EELS).** (a) Plasmon energy expansion thermometry: thermal expansion of crystalline Si due to resistive heating is measured by STEM-EELS using a cold field-emission gun. The inset depicts the image-averaged plasmon energy shift while heating and cooling the Si lamella as compared to the plasmon energy calculated using the temperature-dependent thermal expansion coefficient ($\alpha(T)$). Shaded bars are the plasmon energy standard deviation in each STEM image, and error bars were calculated using reduced $\chi^2$ but smaller than each datapoint. $T_{ref} = 50˚C$. (b) A high-angular annular dark-field STEM image of the membrane at $T_{ref}$, with the Cu grid, 200 nm Ta, and a W capping layer apparent as bright contrast. The dashed red square represents the STEM-EELS acquisition region in (c). Scale bars are 1 μm. False coloration represents hot colors as "warmer" sample regions, and Lorentzian fits were used to determine the plasmon energy. See supplementary movies for temperature-dependent dark-field and plasmon energy images. (d) Temperature-dependent spectra of the Si lamella collected using dualEELS and aligned to the zero-loss peak (ZLP). Each spectrum is an average of spectra in each 40x40 pixel STEM-EELS image with the first two rows used for image alignment. The direct bandgap ($E_g$) and Si $L_{2,3}$ edge are power-law background-subtracted. (e) Calculated Si EEL spectra modeled over a range of expanded lattice parameters using the temperature-dependent, isotropic thermal expansion coefficient of Si [ref. 27]. The unit cell's energy-optimized lattice parameter ($a_0$) was 10.33 Bohr (5.47 Å). A scissor shift was applied to the calculated Si $L_{2,3}$ edge according to the temperature-dependent band gap [ref. 24], and the edge was broadened by an energy-dependent Lorentzian broadening.

# RESULTS

**Figure 1** depicts the experimental setup used to measure PEET and core-loss thermometry. When integrated with STEM-EELS, PEET can accurately map the plasmon energy shift (**Figure 1a**). The measured dual-thickness Si lamella (82.5 ± 2.2 and 106.0 ± 3.5 nm thicknesses) with native oxide was prepared by focused ion beam lift out, thinning, and polishing of >10 kΩ*cm Si (<4.4*$10^{11}$ $cm^{-3}$ intrinsic $N_D$).[26] The inset shows that the Si plasmon energy redshifts linearly with increasing temperature with minor hysteresis across the thermal sweep.

The volume plasmon energy ($E_p$) redshifts due to thermal lattice expansion according to the free electron model in Equation 1:

$$E_p(T) = \hbar\omega_p(T) = \hbar\sqrt{\frac{n(T)e^2}{m_{eff}\epsilon_\infty}} \tag{1}$$

As the Si lattice expands, the carrier density n(T) decreases ($\epsilon_\infty$, e, and $m_{eff}$ are the specimen's dielectric constant, electron charge, and valence band effective mass, respectively). The electron's temperature-dependent effective mass and dielectric constant vary with temperature but are typically ignored by PEET studies by interpreting the *relative* change in the plasmon energy according to the thermal expansion coefficient $\alpha(T)$ using Equation 2:

$$E_p(T) = E_{p0} * (1 - \frac{3}{2} * \int_{T_{ref}}^{T} \alpha(T)dT) \tag{2}$$

Here, $E_{p0}$ is the plasmon energy at the reference temperature ($T_{ref}$) and $\alpha(T)$ is the temperature-dependent thermal expansion coefficient of Si.[27] Equation 2 was used to calculate the temperature-dependent plasmon energy, black trend-line in **Figure 1a**(inset). However, it is apparent that the plasmon shifts 1 meV, or ~25%, more than the predicted value over this temperature range. This shift could be due to a wide variety of factors. The plasmon's absolute energy and redshift can only be directly related to the thermal lattice expansion if approximations for fixed momentum, strain, and dielectric environment across the specimen hold true. We believe that the lamella contains built-in stress from milling and welding with materials of differing thermal expansion coefficients that manifests as temperature-dependent tensile strain. The inset also features two types of error bars (1) capped and (2) shaded. The capped error bars were calculated using a reduced $\chi^2$ analysis of the averaged data points' deviation from a linear fit, normalized to the degrees of freedom but are largely unnoticeable due to the minimal error. The shaded error bars are the standard deviation of the plasmon energy calculated across all the pixels in the STEM-EELS image at each temperature. About half of this ~13 meV standard deviation results from the two different thicknesses of the Si staircase measured in **Figure 1b**. The Si surface-to-volume ratio is thickness dependent because the Si surface plasmon near 10 eV redshifts the volume plasmon with decreasing thickness. The other half of the ~13 meV standard deviation results from suboptimal signal-to-noise ratio in peak fitting. Although each pixel's plasmon signal-to-noise is comparable to other manuscripts in this field, μeV accuracy is needed to properly perform PEET. The experimental PEET uncertainty would improve using higher magnification over a uniform thickness region and dualEELS to separately acquire the ZLP and plasmon spectra.

STEM-EELS (**Figure 1b,c**) was performed on an electron-transparent region of the Si lamella over two different regions of uniform thickness, as indicated by the dashed red square in the high-angle annular dark-field (HAADF) image survey. The 40x40 pixel STEM-EELS image in **Figure 1c** was collected with the sample at $T_{ref}$ = 50 ˚C. The plasmon energy was determined for each pixel where the zero-loss peak (ZLP) was aligned using a center-of-mass fit, three Lorentzian functions were used to fit the Si plasmon peaks, and the energy difference of the ZLP and most intense, first plasmon peak was calculated (see Supplementary Information for fitting details). The thin and thick regions of the specimen are clearly apparent due to the plasmon energy differences. However, the thickness dependence has no noticeable effect on the plasmon energy difference with temperature (see movies in Supporting Information).

During the STEM-EELS mapping, dualEELS spectral acquisition allowed for sequential acquisition of the ZLP, plasmon, multiple scattering, and Si $L_{2,3}$ edge. The spectra averaged across each 40x40 px STEM-EELS image are shown in **Figure 1d**. Collecting the ZLP through dualEELS is highly critical for core-loss thermometry in a (S)TEM as it allows for longer acquisitions of the low-intensity core-loss spectrum while still offering a method for spectral alignment. The spectral redshift with temperature is apparent in the experimental and calculated plasmon and $L_{2,3}$ edge spectra (insets in **Figure 1d,e**).

The direct bandgap ($E_{g,direct}$) of Si, Si main plasmon, and Si $L_{2,3}$ edge are highlighted with insets showing the magnified features' redshifts. The Si surface plasmon around 10 eV is ignored, as well as the $SiO_2$ plasmon that is not resolvable at this thickness. Temperature-dependent spectra in **Figure 1e** were modeled using TDDFT for the ~16.8 eV volume plasmon (turboEELS) and DFT/BSE for the ~99 eV Si $L_{2,3}$ edge (OCEAN). Thermal lattice expansion was modeled using lattice parameters spanning $a/a_0$ = 1.000 – 1.002 according to the temperature-dependent thermal expansion coefficient.[27] The lattice parameter was the only input parameter modified across simulations, and the band gap was corrected after the calculation to account for the well-known band gap underestimation using DFT.[28] The reference lattice parameter ($a_0$ = 10.33 Bohr) was selected by minimizing the total cell energy.

The *ab initio* TDDFT calculations match the analytical approach for PEET as apparent in the **Figure 1e** insets. For PEET on Si, *ab initio* calculations are generally not needed as the thermal lattice expansion is well-studied.[6] However, more complex materials like transition metal oxides and multi-material stacks/interfaces that contain multiple plasmon and low-loss features are challenging to deconvolute. Instead, an *ab initio* TDDFT approach like the one used here would have to be applied.[7] Core-loss thermometry is therefore motivated because its element-specificity leads to well-separated spectral features in complex materials. However, we first must test whether a simple analytical procedure can sufficiently predict the temperature-dependent core-loss spectra without relying on an *ab initio* approach.

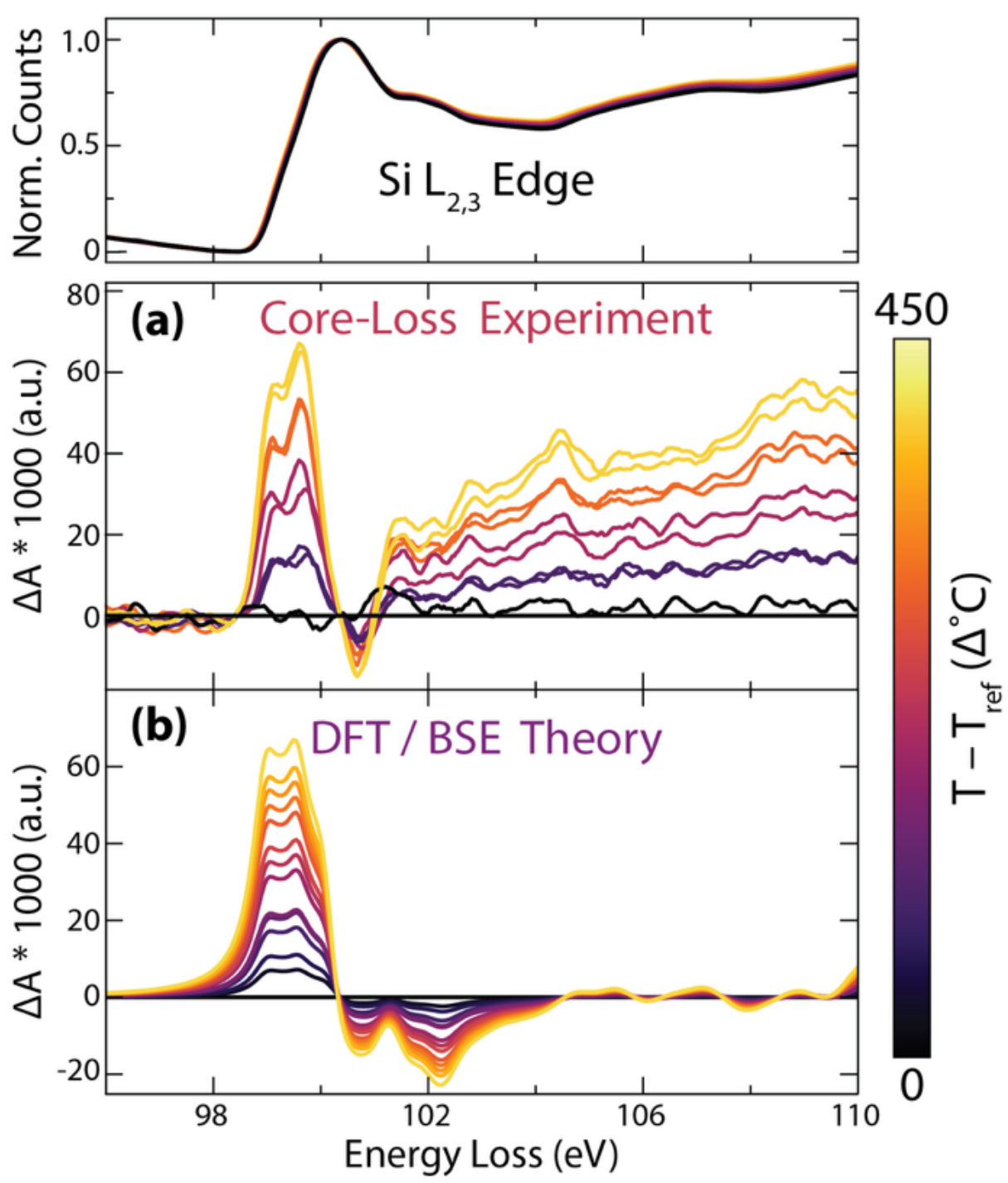

**Figure 2. Si $L_{2,3}$ core-loss differential EEL spectra.** Measured (a) and modeled (b) differential spectra of the Si $L_{2,3}$ edge from Figure 1d,e. Spectra are calculated as ($A_T$ – $A_{Tref}$) * 1000. The measured differential is smoothed by a Savitsky-Golay filter (window = 17, order = 3). Two measurements were taken at each temperature, one for both the cooling and subsequent cooling ramps in the Figure 1a, inset. Modeled spectra were calculated using *ab initio* DFT / BSE theory with a scissor shift applied according to the temperature-dependent band gap [ref. 24]. The as-measured temperature-dependent Si $L_{2,3}$ edge spectra are shown above for reference, and the measured spectra exclude any pixels over the Ta alignment layer.

Conceptually, the interpretation of temperature-dependent core-level spectra is complicated by complex dependencies of the specimen's band gap, band curvature, and dielectric function following thermal lattice expansion, lattice vibrational modes, and strain; thus, the mechanism for the previously reported redshift of core-level spectra with temperature remains unclear.[21–23] To help discern the origin of the redshift, we measure the temperature-dependent differential spectra and compare these results to calculated spectra using a DFT/BSE approach (**Figure 2**). The core-loss spectra were aligned for each STEM-EELS pixel using the center-of-mass-fit ZLP and averaged across each STEM-EELS map. This approach assumes that drift tube hysteresis does not occur tube at each pixel. The differential spectra were calculated by first normalizing each core-loss spectrum to span zero to one for the core-loss near edge's minimum and maximum. The spectrum at $T_{ref}$ = 50 ˚C = ~325 K was then subtracted from the corresponding spectrum at each elevated temperature and multiplied by 1000 for visualization (**Figure 2a**). The differential spectra for the DFT/BSE-calculated spectra were similarly obtained (**Figure 2b**).

The differential spectra in **Figure 2** have positive features at the edge onset, indicating the redshift of the core-loss spectra with temperature and lattice expansion as apparent in the **Figure 1d,e** insets. The positive differential feature is almost identical to a rigid spectral shift (**Figure S8**), which suggests the conduction band's density of states is minimally modified at the measured temperatures. X-ray studies often normalize to an intensity far past the edge onset, but multiple scattering and temperature-dependent constructive interference effects in EELS prevented this normalization method.[29] These effects appear as increasing intensity with temperature beyond the edge (>101 eV). The DFT/BSE approach used here is also optimized for near-edge features instead of post-edge effects.

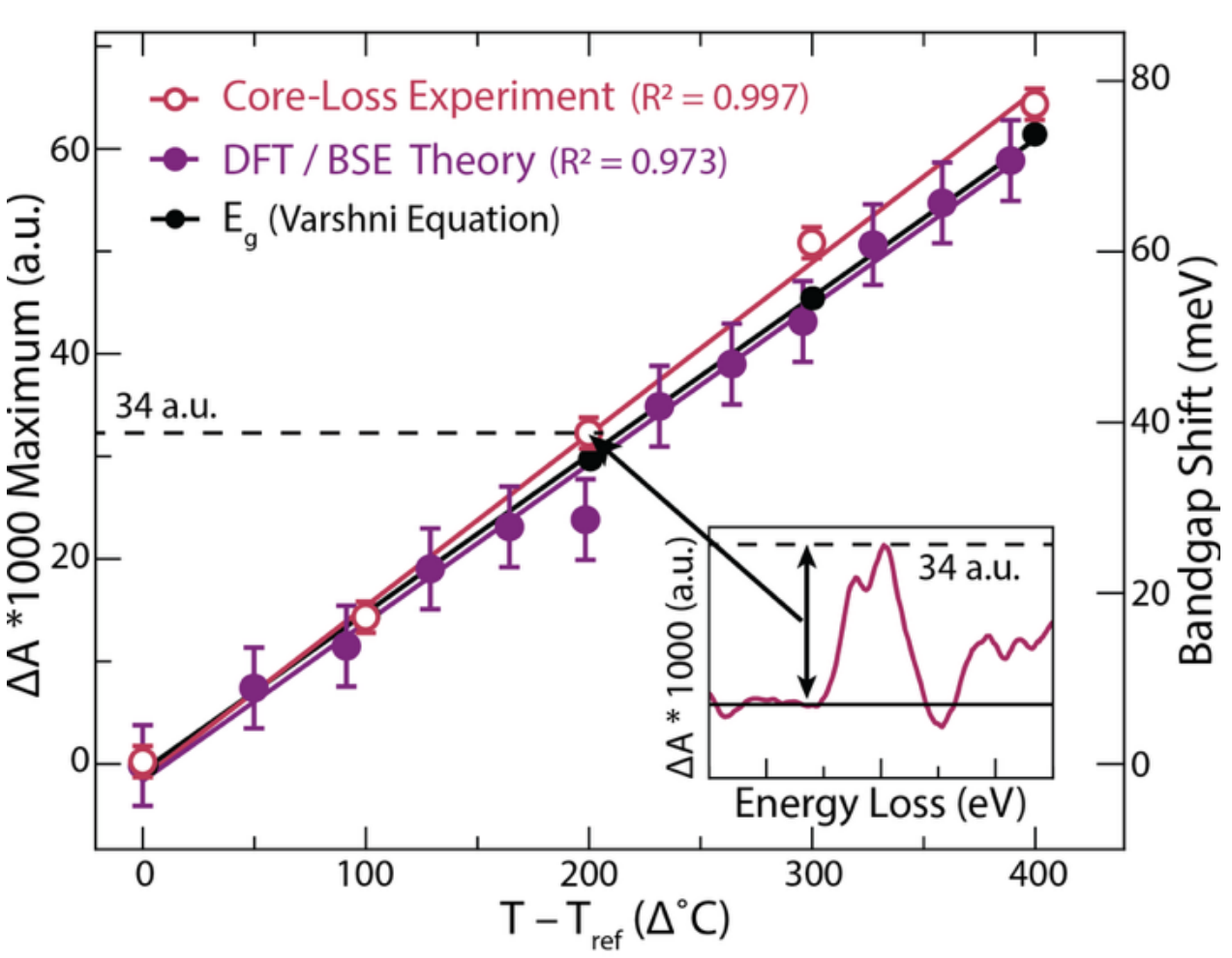

**Figure 3. Comparing the modeled and measured Si $L_{2,3}$ edge's differential amplitude and redshift.** The average maximum differential amplitude of the measured (magenta open circles) and simulated (purple filled circles) core-loss differential spectra from Figure 2. On the right axis, the amplitude is converted to the band gap's redshift according to a rigid shift model. The temperature-dependent band gap of Si from the Varshni equation is shown for reference (black filled circles). $T_{ref}$ = 50 ˚C. The inset shows an example differential spectrum where the maximum amplitude was measured.

The maximum differential intensity is calculated to graph the relationship between the temperature-dependent band gap shift for the core-loss experiment and theory (**Figure 3**). To convert the differential intensity into the energy of the edge's redshift, a rigid shift of the spectrum at $T_{ref}$ is performed (5 meV shift steps up to 150 meV), and linear regression is used to convert the differential intensity into the meV of the redshift. Both the measured and calculated core-loss spectra are found to follow the Varshni equation. A Varshni equation is an empirical formula that predicts a semiconductor's band gap ($E_g$) as a function of temperature (T).[24,25] A Varshni equation typically has the form $E_g = E_0 – (a * T^2) / (T + b)$ where a and b are derived constants and $E_0$ is the band gap at 0 K. Whether experimentally measured or computationally derived, an accurate Varshni equation will take into consideration both the thermal lattice expansion and lattice vibrations from electron-phonon renormalization. For covalently bonded materials like Si, the electron-phonon vibrational term dominates because the lattice only minimally expands.

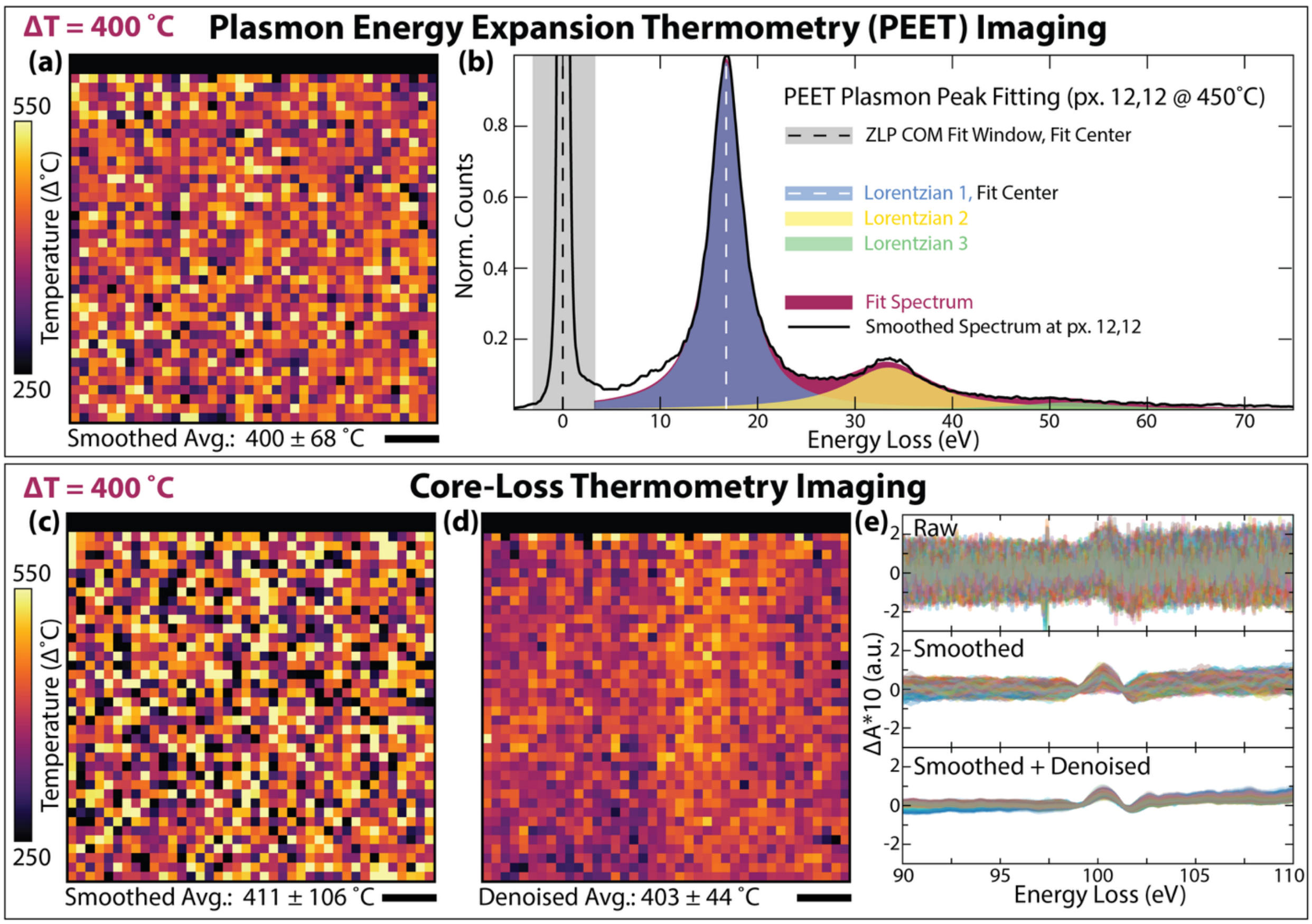

**Figure 4. Comparing the accuracy of plasmon energy expansion and core-loss thermometry imaging.** (a) A plasmon energy expansion thermometry (PEET) image of the Si lamella for ΔT = 400 ˚C. (b) Each pixel's spectrum was smoothed and aligned using a center-of-mass (COM) fit of the zero-loss peak (ZLP). The Si plasmon energy was calculated in the 450 ˚C and $T_{ref}$ images, and each pixel's plasmon energy difference between 450 ˚C and $T_{ref}$ was converted to a measured temperature using linear regression from **Figure 1a**(inset). (c) A core-loss thermometry image of the Si lamella for ΔT = 400 ˚C. Each pixel's core-loss spectrum was again aligned using the ZLP COM, and a differential spectrum was calculated and smoothed for each pixel. The amplitude of the differential peak resulting from bandgap reduction was converted to temperature using linear regression from **Figure 3**. (d) A core-loss thermometry image denoised using a 3-layer bottleneck autoencoder from scikit-learn's MLPRegressor. (e) The denoiser uses self-supervised learning by leveraging the smoothed spectral output trained on 10% of the data. All scale bars below the STEM-EELS images are 250 nm.

**Figure 3** verifies that the core-loss redshift magnitude can be used to determine the Si lattice temperature according to the Varshni equation. Similarly, core-loss thermometry could enable a new approach to directly determine the constants in the Varshni equation for semiconductors with sub-nm resolution. By determining the constants for bulk specimens using simpler methods (e.g., quantifying the bandgap with UV-Vis spectroscopy), core-loss thermometry could then be used to derive the Varshni equation's empirical constants in compositionally complex materials without relying on DFT/BSE calculations.

The novelty of performing thermometry in the STEM is the ability to measure temperature with nanoscale precision, and thus, the accuracy of core-loss thermometry imaging must be tested. **Figure 4** compares the difference images for ΔT = 400 ˚C. Our PEET imaging result (**Figure 4a**) indicates that the average temperature is measurable at 400 ± 68 ˚C, 17% error. This error is the standard deviation across the difference image, and it indicates that dualEELS is also necessary to improve the signal-to-noise ratio of plasmon spectra for accurate PEET. Notably, this error could be minimized by increasing the magnification and acquisition time, but the larger relative sample drift and beam-induced heating should still be considered. Minimal error resulted from the two different specimen thicknesses at this level of accuracy, which has been tested in the literature.[13] Each pixel was analyzed using Lorentzian fits of the Si plasmons, where the main plasmon's Lorentzian fit center at each temperature is a proxy for lattice temperature (**Figure 4b**). The PEET image is generated by calculating the difference in the plasmon energy between $T_{ref}$ and T = 450 ˚C at each pixel ($\Delta T$ = 400 ˚C), and the plasmon energy difference is converted to temperature using linear regression from **Figure 1a**, inset.

Core-loss thermometry imaging was similarly performed for each pixel but instead analyzed using the Si $L_{2,3}$ edge differential intensity as in **Figure 2a** and linear regression from **Figure 3**. Each spectrum in the image is smoothed using a Savitsky-Golay third-order smoothing function over 51-spectral pixels (**Figure 4c**). The core-loss thermometry image's average temperature with spectral smoothing considered is 411 ± 106 ˚C (26% error). This relatively high standard deviation is a result of the poor signal-to-noise ratio at each pixel. Therefore, by applying a standard denoising function, trained by 10% of the STEM-EELS core-loss spectra, we further optimize the differential signal while minimizing the background noise (**Figure 4d**). The denoised core-loss spectral image obtains a measured temperature of 403 ± 44 ˚C (11% error). This 58% reduction in absolute uncertainty from the non-denoised core-loss thermal image can be further optimized for samples of uniform thickness or more robust denoising algorithms.

# DISCUSSION

**Table 1. Thermal lattice expansion calculation methods in this work.** All measurements were performed on crystalline Si with native oxide. The reported temperature-dependent thermal expansion coefficient (TEC) of Si was used for all inputs [ref. 27], and its fit error represents the nonlinearity in this temperature range of 50 – 450 ˚C.

| TEC Measurement | Method | Required Inputs | TEC [$10^{-6}$ $K^{-1}$] | Fit Error*100 (and $R^2$) |
|---|---|---|---|---|
| Reported TEC [ref. 27] | – | – | 4.09 | 0.37 (0.996) |
| Plasmon thermometry | Low-loss EELS | Free electron model ($\epsilon_\infty$, $m_{eff}$ ) | 5.06 | 0.32 (0.998) |
| Plasmon thermometry | TDDFT | Free electron model ($\epsilon_\infty$, $m_{eff}$ ), TEC | 3.51 | 0.41 (0.994) |
| Core-loss thermometry | Core-loss EELS | Varshni equation[a], TEC | 3.89 | 0.52 (0.993) |
| Core-loss thermometry | DFT/BSE | Varshni equation[a], TEC | 3.98 | 0.59 (0.989) |

[a] The Varshni equation's constants can be taken from literature or derived using the core-loss spectral shift

**Figure 5** and **Table 1** compare the thermal expansion coefficients derived using PEET and core-loss thermometry to literature-reported values of crystalline Si. In PEET, the plasmon energy shift is converted to lattice expansion using the free-electron model (**Equation 1**) as: $a(T) / a(T_{ref})$ =

$(E_p(T) / E_p(T_{ref}))^{2/3}$, where a is the lattice parameter. This approach assumes isotropic expansion. To derive the lattice parameter change from core-loss thermometry, we convert the as-measured Si $L_{2,3}$ and bandgap redshift with temperature by using the reported Varshni equation for Si.[24]

PEET overestimates the thermal expansion coefficient by $1.03*10^{-6}$ $K^{-1}$, approximately 25% higher than the literature-reported value. This discrepancy may arise from temperature-dependence of the dielectric constant, valence band effective mass, sub-pixel specimen drift, or lamella strain. While the effective mass is known to increase by ~5-10% over this temperature range, a substantial shift in the dielectric constant would also be required to fully explain the observed error.[30,31] In contrast, core-loss thermometry provides a more accurate measure of the thermal expansion coefficient at $3.89*10^{-6}$ $K^{-1}$, with reduced uncertainty in the image-averaged dataset even without spectral denoising. TDDFT and DFT/BSE simulations, based on isotropic lattice expansion, show relatively good agreement with both respective thermometry methods.

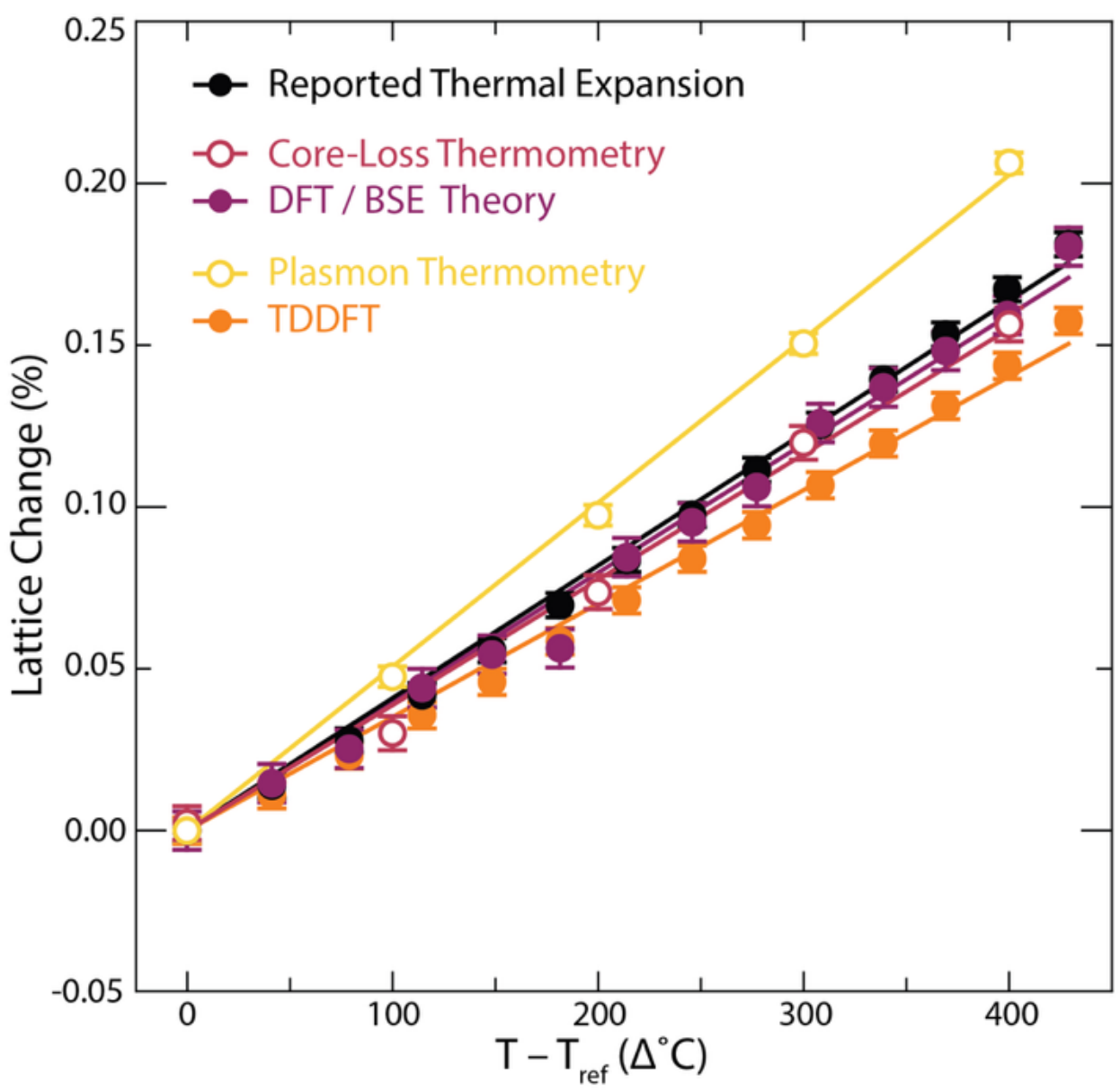

**Figure 5. Thermal lattice expansion quantified using low- and core-loss EELS.** Four methods are compared against the reported thermal expansion of crystalline Si [ref. 27]. Core-loss thermometry (magenta) quantifies lattice expansion via shifts in the Si $L_{2,3}$ edge arising from the temperature-dependent band gap, while plasmon thermometry (yellow) tracks the plasmon energy shift according to the free-electron model. DFT/BSE (core-loss) and TDDFT (plasmon) simulations were performed by isotropically expanding the Si unit cell; DFT/BSE spectra were corrected for the intrinsic bandgap. The slope of each fit corresponds to the thermal expansion coefficient (see Table 1), with y-intercepts fixed at 0%. Error bars are determined from reduced $\chi^2$ analysis of the linear fits.

One primary drawback of core-loss thermometry is that its temperature-dependence arises from more complex physical phenomena than the direct lattice expansion captured by PEET. For example, the thermal core-loss change would likely be more subtle for metals without a band gap. PEET is best suited for measuring thermal expansion of pure, uniform metals with a single, well-defined plasmon peak. As such, the two methods are highly complementary. This study introduces core-loss thermometry as a new approach for semiconductors and still requires validation for other materials with complex temperature-dependent changes in the electronic structure.

Other practical factors of this experiment must be considered. While the intrinsic defects or native $SiO_2$ layer could potentially introduce local variations in the thermal expansion, their concentrations were negligible ($4.4*10^{11}$ / $5*10^{22}$; 4 nm / 90 nm = 4% $SiO_2$ concentration).[26,32] Potential phase transitions and contamination were avoided by specimen annealing, and the immeasurable oxygen K-edge further confirms the insignificant $SiO_2$ concentration. Future studies could quantify temperature-dependent strain, tilt, and momentum effects on the accuracy of PEET and core-loss thermometry.[33] To investigate any microscope- and sample-induced STEM-EELS artifacts, we tested correlations between the dark-field, PEET, plasmon energy, core-loss thermometry, and thickness images across the 40x40 pixel scan area. The results show a strong statistically significant correlation between thickness and plasmon energy (0.733, p-value = ~0;

apparent in **Figure 1c**) and a weak yet significant correlation between the denoised core-loss temperature map and thickness (0.226, p-value = ~0; apparent in **Figure 4d**). See Supplementary Information for all correlations. PEET datasets show no correlations. Balancing the beam current, length scale, and pixel view time will be important to approach sub-nm spatial resolutions without inducing beam damage and heating (see Supplementary Information). High-stability MEMS *in situ* heating holders will allow enhanced stability needed for atomic-scale STEM-EELS thermometry.

To improve pixel-level precision in core-loss thermometry, we implemented an open-source, self-supervised autoencoder to denoise the differential core-loss spectra. Denoising is an emerging approach to electron microscopy, enabling faster acquisition times of features even beyond existing resolution limits.[34–37] Our autoencoder was trained on 10% of the data to reconstruct spectra by distinguishing intrinsic signals from noise. This approach reduced uncorrelated pixel-level noise by 58% while preserving the shape and position of spectral features, far more accurate than smoothing alone. However, the specimen's thickness variability posed challenges for model training. Models trained on mismatched thicknesses introduced noticeable artifacts due to differences in the plasmons' multiple scattering background. This artifact is apparent in the denoised core-loss thermometry image in **Figure 4d**, where thicker regions appear as brighter contrast (hotter temperatures) due to the rising background. Therefore, optimal denoising occurs when the training set's background matches that of the test set. Additionally, smoothing spectra prior to denoising improved model performance by supressing the high-requency noise poorly modeled by this network. Despite these considerations, our denoising strategy is a promising step toward reliable, high-resolution core-loss thermometry. Continued optimization of data acquisition and denoising model robustness will be essential for advancing core-loss thermometry's accuracy.

Through a combination of *in situ* STEM-EELS and first-principles simulations, we demonstrate core-loss thermometry for imaging element-specific lattice temperatures in semiconductors with nanometer-scale precision. By directly probing temperature-dependent changes in electronic structure, core-loss thermometry is used to capture the redshift of the Si $L_{2,3}$ edge – an effect that tracks the bandgap's redshift as driven by thermal lattice expansion and electron-phonon renormalization. As a result, this technique not only enables temperature mapping but also extracts Varshni equation constants with sub-nanometer precision. Despite the inherent complexity of core-loss EELS, our findings show that core-loss thermometry yields an interpretable signal that does not rely on *ab initio* calculations, making the approach broadly accessible to the X-ray spectroscopy and electron microscopy communities. Compared to PEET, core-loss thermometry offers improved accuracy for semiconductors with low thermal expansion coefficients or non-uniform composition and strain. With continued development in spectral alignment, denoising, and acquisition, core-loss thermometry will map temperatures at sub-nanometer spatial resolutions during operando or ultrafast experiments. We envision future investigations of thermal strain and heat transfer in multi-element junctions and at nanoscale interfaces relevant to microelectronics, photonics, and photocatalysts.

## METHODS

### Temperature-dependent STEM-EELS

All TEM measurements were performed on a JEOL NeoARM microscope with a cold field-emission gun (zero-loss peak FWHM = 0.6 eV). STEM-EELS data was acquired at 200 kV in 40kx STEM magnification mode using a 3 cm camera length. The sample was measured at ~0˚ tilt, approximately on the [100] zone axis. The tilt angle was not controlled during the thermal sweep. The beam current was approximately 100 pA or 55,800 pA/μm$^2$ in STEM. The Gatan Continuum Imaging Filter was aligned using a 2.5 mm entrance aperture with 30 meV/ch dispersion and a K3 detector. The Gatan sub-scan feature (32x32/pixel) was enabled to diffuse the beam over the scanning region on the sample and minimize beam-induced heating. The acquisition time was fixed at a 0.25 s view time in dualEELS (1% live time low-loss with 0 eV shift, 100% live time high-loss with 20 eV shift). Specimen drift was corrected every 40 pixels and managed using Gatan drift-correction software over three image-averaged passes. The specimen drift was 10 – 20 % of the 42.33 nm pixel size between each drift correction step.

The Si lamella was prepared from a 200 nm Ta-capped, 100-oriented intrinsic Si wafer substrate (>10 kΩ*cm), University Wafer. The autoTEM lift-out feature on a ThermoFisher Helios Ga FIB was utilized to lift-out the specimen with a W capping layer deposited on the sample via ion-beam induced deposition (IBID) at 30 kV prior to lift-out. The lamella was mounted on a Cu lift-out grid. It was thinned into the staircase shape using 30 kV ions. Accelerating voltages of 5 kV and 2 kV were subsequently used for final thinning and polishing. A Gatan 652-TA model heating holder was used for all temperature-dependent STEM-EELS measurements. The specimen was first loaded into a vacuum pumping station and heated to 450 ˚C for 30 minutes to anneal and minimize surface carbon contamination, and it was again heated to 450 ˚C for 30 minutes inside the microscope column to minimize transfer-induced contamination. During STEM-EELS measurements, the holder temperature was first set to 450 ˚C and the temperature was ramped down in 100 ˚C steps to 50 ˚C and then ramped up to check for heating hysteresis. Between each temperature set point, 5 minutes was allotted for thermal equilibration. Gatan reports a ±5 ˚C accuracy for the heating controller.

All STEM-EELS data was analyzed using the HyperSpy python loader.[38] A 19 px-wide median filter was applied to all data to remove hot pixels. Center-of-mass fitting was used for ZLP peak fitting, and three Lorentzian functions were used to fit the three Si plasmons. Additional fitting and spectral denoising parameters are detailed in the Supplementary Information. All analysis scripts can be provided upon request.

## *Ab Initio* Theory

**Density Functional Theory (DFT) Parameters.** A 200 Ry cutoff energy was used for all DFT calculations based on convergence tests with a 10.33 Bohr lattice parameter (see test results in Supplementary Information). All DFT self-consistent field calculations utilize a $1\text{x}10^{-10}$ Ry convergence threshold, 16x16x16 *k*-point mesh, 6 electronic states (bands), and standard atomic positions for crystalline Si in the diamond cubic crystal structure (fcc lattice with a two-atom basis). The silicon pseudopotential file for all calculations was produced using the ONCVPSP code.[39] This norm-conserving pseudo was calculated using a PBE functional and nonlinear core correction.

**Low-Loss EELS Theory.** turboEELS was used to simulate the Si volume plasmon, a program integrated within the Quantum ESPRESSO package.[40,41] turboEELS calculates the low-loss volume plasmon spectra using a Liouville–Lanczos approach to linearized time-dependent DFT (TDDFT). By calculating the charge-density susceptibility with the quantum Liouville equation, the inverse of the imaginary/longitudinal component of the dielectric function (the EEL spectrum) is simulated. EEL spectra were calculated using DFT and BSE theory.

**Core-Loss EELS Theory.** To simulate the Si $L_{2,3}$ edge core-loss spectrum, we leverage the Obtaining Core Excitations from the *Ab initio* electronic structure and the NIST BSE solver (OCEAN) package.[42] The OCEAN workflow uses Quantum ESPRESSO to perform DFT calculations of silicon's electronic structure and subsequently calculates the core-hole exciton effects and core-level spectrum through BSE calculations. OCEAN 3.0.1 and Quantum ESPRESSO 7.0 were used for all calculations. Due to DFT's well-known inaccuracy in calculating the band gap of materials, a scissor shift corresponding to the temperature-dependent band gap is incorporated for each lattice parameter simulated. The BSE and non-self-consistent field and screening calculations simulate 100 electronic states (bands). The static dielectric constant is fixed to $\varepsilon = 11.4$.[43]

## ACKNOWLEDGEMENTS

The authors thank Professor Ye-Jin Kim, Professor Oh-Hoon Kwon, Dr. Haihua Liu, Nicholas Heller, and Nicholas Hagopian for helpful research discussions regarding project directions and data analysis. The authors also thank Dr. Jianguo Wen for contributing to sample preparation and microscope alignment.

This research was supported as part of the Ensembles of Photosynthetic Nanoreactors, an Energy Frontier Research Center funded by the U.S. Department of Energy, Office of Science under Award No. DE-SC0023431. Work performed at the Center for Nanoscale Materials, a U.S. Department of Energy Office of Science User Facility, was supported by the U.S. DOE, Office of Basic Energy Sciences, under Contract No. DE-AC02-06CH11357. Sandia National Laboratories is a multi-mission laboratory managed and operated by National Technology and Engineering Solutions of Sandia, LLC., a wholly owned subsidiary of Honeywell International, Inc., for the U.S. Department of Energy's National Nuclear Security Administration under contract DE-NA0003525. The views expressed in this article do not necessarily represent the views of the U.S.

Department of Energy, National Science Foundation, or the United States Government. The computations presented here were conducted in the Resnick High Performance Computing Center, a Resnick Sustainability Institute facility at the California Institute of Technology.

L.D.P. was supported by the National Science Foundation Graduate Research Fellowship under Grant No. DGE-1745301. This material is based upon work supported by the U.S. Department of Energy, Office of Science, Office of Workforce Development for Teachers and Scientists, Office of Science Graduate Student Research (SCGSR) program. The SCGSR program is administered by the Oak Ridge Institute for Science and Education for the DOE under contract number DE-SC0014664. W.L. acknowledges support from the Korea Foundation for Advanced Studies.

**AUTHOR INFORMATION**

**Affiliations**

[1] *Division of Chemistry and Chemical Engineering, California Institute of Technology, Pasadena, CA 91125, USA*
[2] *Center for Nanoscale Materials, Argonne National Laboratory, Lemont, IL 60439, USA*
[3] *Sandia National Laboratories, Livermore, California 94551, USA*

**Contributions**

L.D.P., W.L., T.E.G., and S.K.C. conceptualized and designed the project scope. L.D.P. acquired the data, performed the data analysis, and prepared the manuscript. L.D.P. and W.L. performed the computational modeling. T.E.G. led the microscopy efforts, approach, and alignment. S.K.C supervised the project and its funding. All authors provided feedback and direction during the manuscript submission and review.

**Corresponding Author**

Correspondence to Thomas E. Gage & Scott K. Cushing

**TOC Graphic**

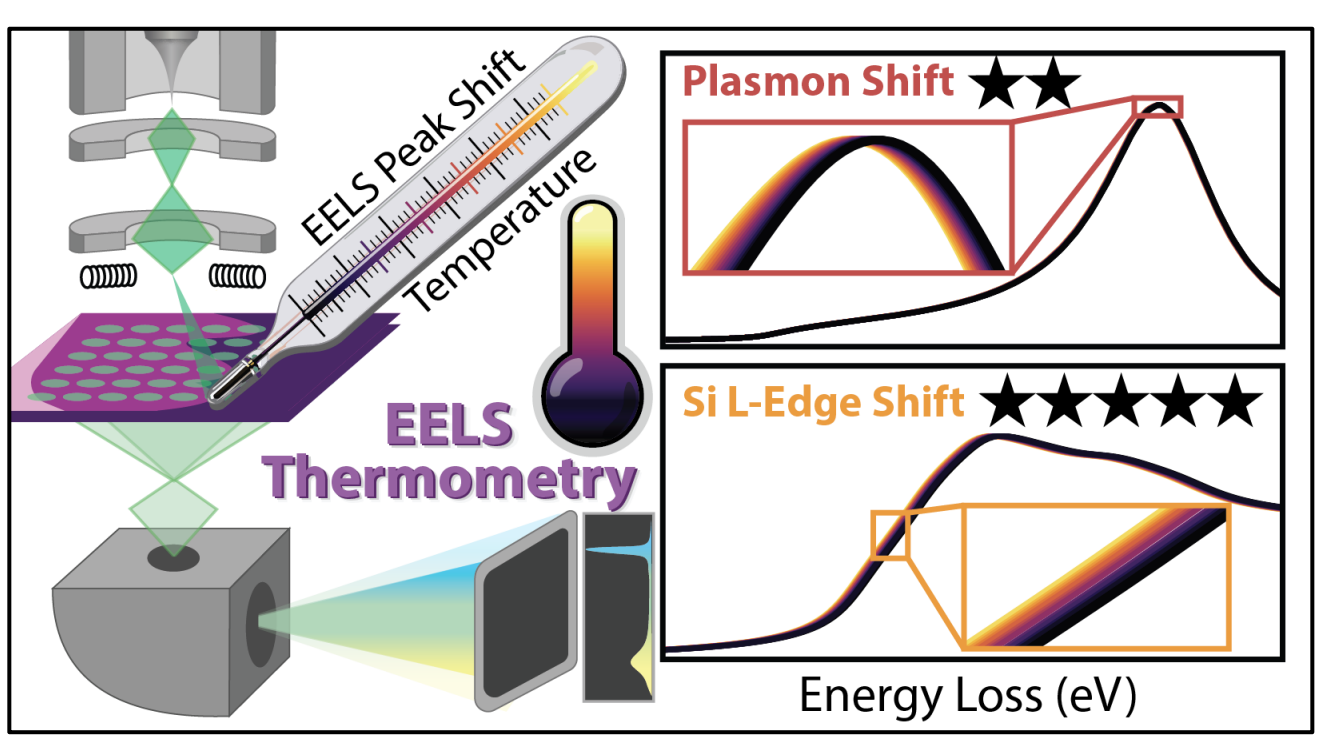

# SUPPLEMENTARY INFORMATION

# Nanoscale and Element-Specific Lattice Temperature Measurements using Core-Loss Electron Energy-Loss Spectroscopy

Levi D. Palmer,[1] Wonseok Lee,[1] Daniel B. Durham,[2] Javier Fajardo, Jr.,[3] Yuzi Liu,[2] A. Alec Talin,[3] Thomas E. Gage,[2,*] Scott K. Cushing[1,*]

*[1]Division of Chemistry and Chemical Engineering, California Institute of Technology, Pasadena, CA 91125, USA*
*[2]Center for Nanoscale Materials, Argonne National Laboratory, Lemont, IL 60439, USA*
*[3]Sandia National Laboratories, Livermore, California 94551, USA*

*Correspondence to Thomas E. Gage & Scott K. Cushing

## Further Analysis

### Beam-Induced Heating

Sample heating poses further challenges for accurate core-loss thermometry when using a high electron beam current to optimize the signal-to-noise or very high magnifications. A microscope's beam current is typically increased to adequately resolve core-loss edges, especially those beyond 1 keV. In this work, the beam current and density used were around 100 pA or 55,800 pA/μm$^2$ for the 42x42 nm pixel size. The beam current decreased throughout each down/up temperature sweep by ~15% due to the cold FEG emission decay, but we will interpret the specimen heating from the maximum beam current for best practice.

Following analysis from prior PEET literature [ref. 6, main text], 200,000 V * 100*10$^{-12}$ A = 20 μW of beam power. Roughly 75% of the electrons at this voltage lose energy to the sample, so 15 μW are deposited at ~16.8 eV/e$^-$. Therefore, the beam input is 15 μW * (16.8 eV / 200 keV) = 1.26 nW over the 1,789 nm$^2$ STEM-EELS pixel area.

Considering the 1 W/(m*K) thermal barrier of the ~5 nm thick Si oxide, the beam-induced thermal shift would be ~0.01 K per pixel, far below both techniques' resolution limits. The thermal resistance (R) of the $SiO_2$ layer is calculated as:

R = thickness / (thermal conductivity * area) = 5*10$^{-9}$ / (1 * (0.00179*10$^{-12}$)) = 2.8*10$^6$ K/W

Where the expected temperature increase can be calculated as:

$$\Delta T = P * R = (1.26*10^{-9}\ W) * (2.8*10^{6}\ K/W) = 0.0035\ K$$

Although this calculation suggests that beam-induced sample heating does not influence thermometry in this work, a careful balance of beam current and dwell time will be essential for mitigating beam-induced artifacts while still acquiring the signal-to-noise needed for core-loss thermometry at sub-nm length scales. This is especially true for thermal measurement accuracy within ±1 K.

### Denoising STEM-EELS Differential Spectra

A denoising autoencoder was implemented to reduce pixel-level noise in the differential low-loss spectra. A multilayer perceptron regressor (MLPRegressor, scikit-learn) was trained on the 14 – 118 eV spectral range using all spectra in the dataset except for the top two rows, which were omitted due to known detector artifacts. The model architecture included three hidden layers (32–8–32 neurons) with ReLU activation. Spectra were flattened and randomly split into training and test sets (90%/10%), and the model was trained to reproduce its input (self-supervised). The trained model was then used to denoise the entire dataset. This method preserved the physical peak shape and energy position while suppressing uncorrelated noise across the spatial domain. All denoising was performed in post-processing and did not affect fitting procedures or raw spectra interpretation.

## Experimental Data

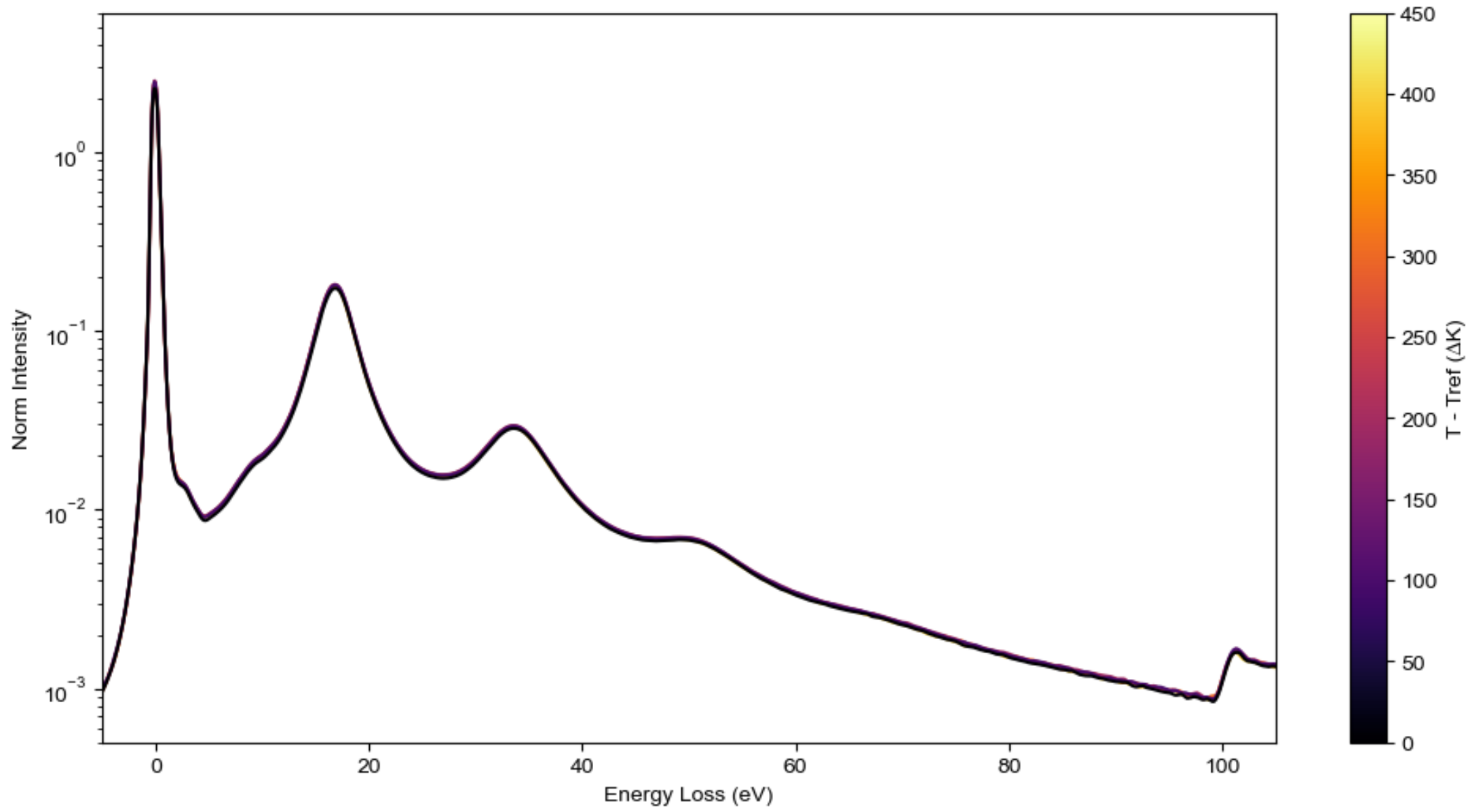

**Figure S1. The entire low-loss EEL spectrum in log scale.** All low-energy spectra are plotted together in log scale without the dualEELS core-loss (high-loss) spectra displayed. By using longer acquisition times and without spectral drift, the core-loss Si $L_{2,3}$ edge would be resolvable at this dispersion.

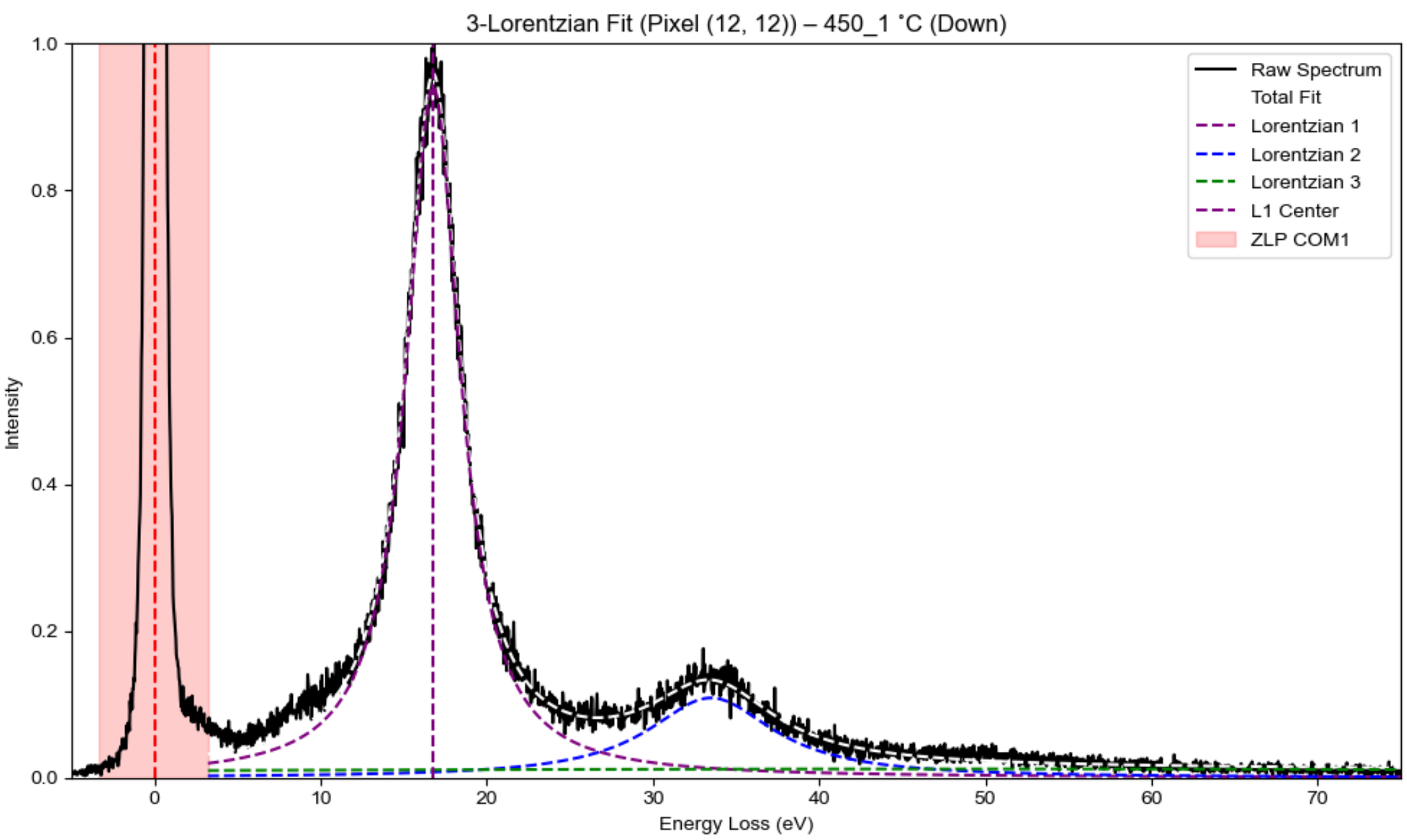

**Figure S2. Visual inspection of the low-loss spectral fitting for unsmoothed data.** Smoothing was only incorporated to improve the statistics for the PEET map in **Figure 4a**, main text. Smoothed spectra used for PEET imaging are plotted in **Figure 4b**, main text, which includes a 19-px median filter for hot pixels and a 17-px, 2[nd] order Savitsky-Golay filter for smoothing.

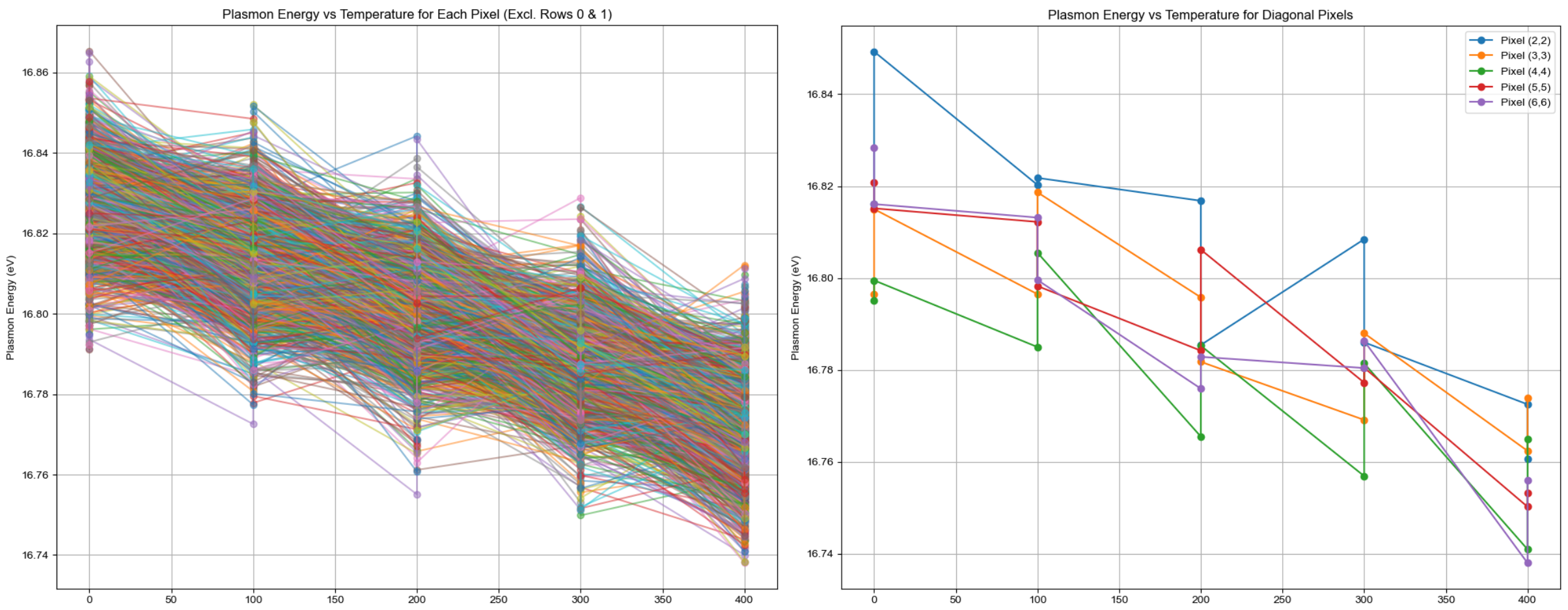

**Figure S3. Plasmon energy variations at individual pixels and across temperatures.** To compare the effects of the STEM scan's pixel-dependent beam alignment and the Lorentzian fit accuracy, the plasmon energy at each pixel across all measured temperatures is displayed. The left image displays all pixels (1520), and the right image displays only the first five pixels across the image diagonal. The first two rows in the STEM-EELS image are excluded as they are the Ta alignment layer.

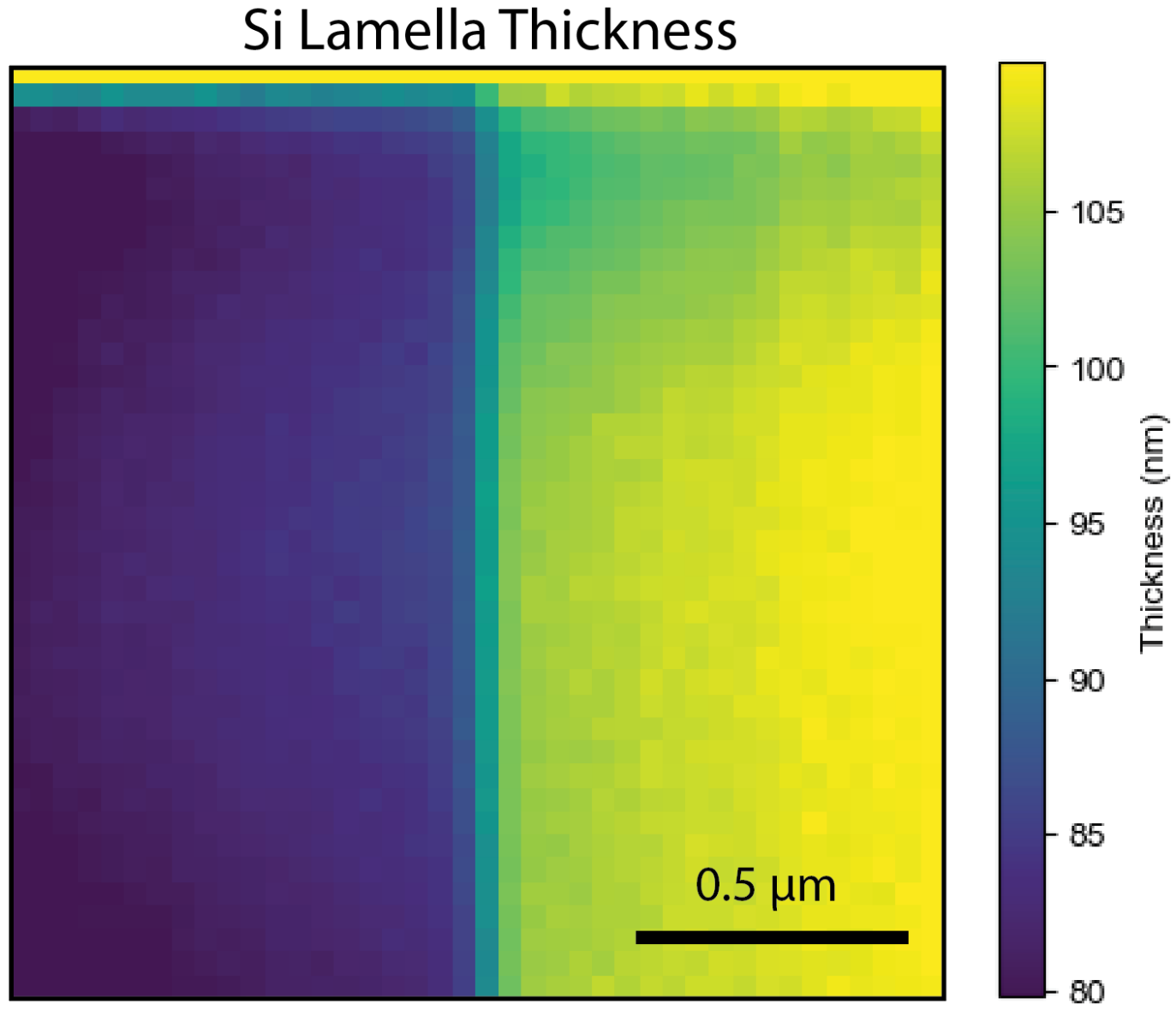

**Figure S4. Thickness mapping across the STEM-EELS low-loss image at $T_{ref}$.** The log-ratio method was used to calculate the sample thickness by summing the electron counts over the ZLP (300 px, 9.8 eV window) and the plasmon region (3150 px, 103 eV window). The reported mean-free path for Si (160 nm) was used [ref. 26, main text]. Average thickness = 94±12 nm; left side thickness = 83 ± 2 nm; right side thickness = 106 ± 4 nm. The full HAADF image can be found in **Figure 1b**.

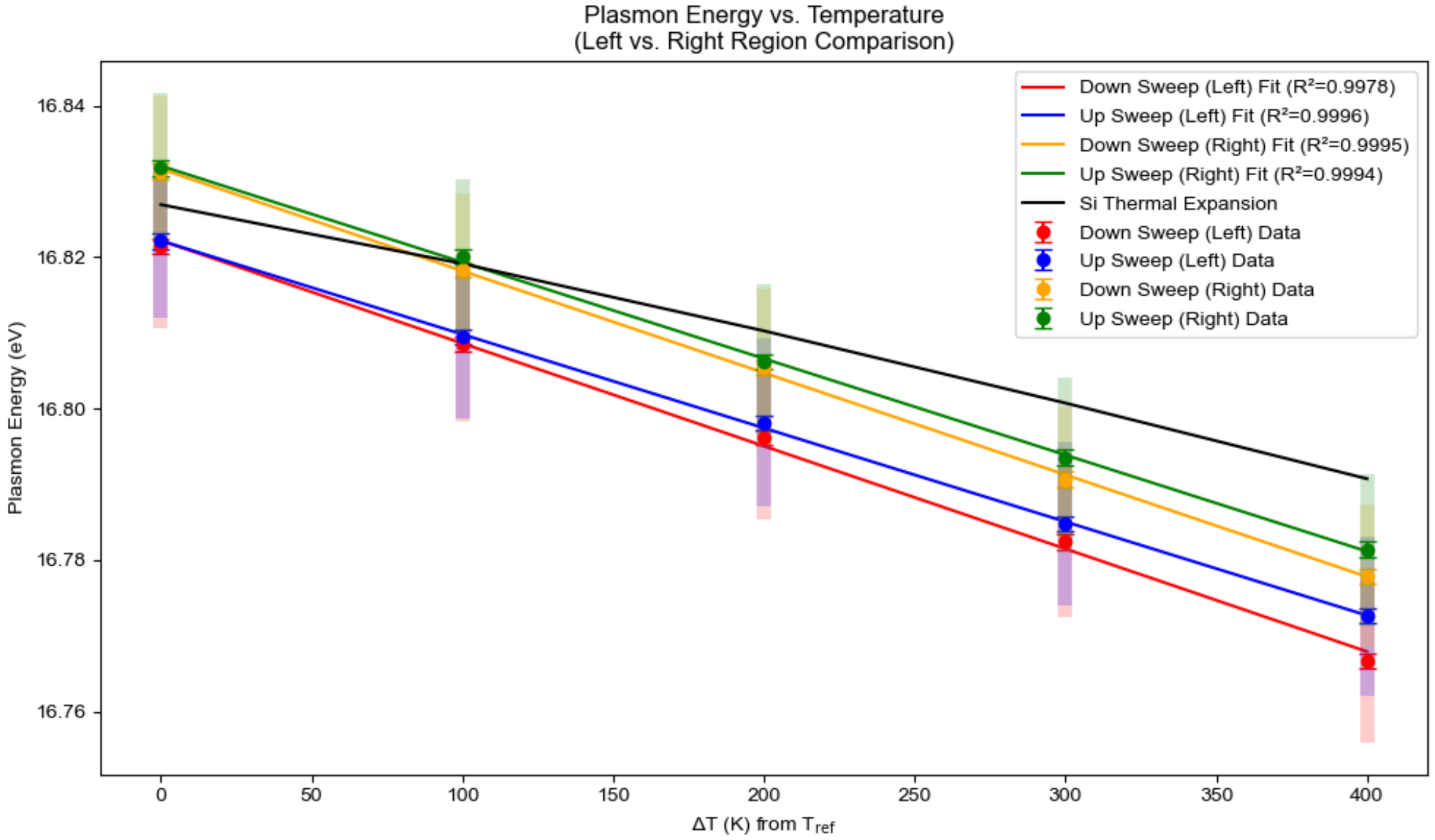

**Figure S5. Plasmon Thermometry on Different Thicknesses.** The plasmon energy varies as a function of sample thickness, likely due to the surface-to-volume ratio difference that causes the Si surface plasmon to redshift the Lorentzian fit center. Here, the trends represent the linear fits of the plasmon energy shift for the thin/left side (red and blue) and thick/right side (green and yellow) of the specimen. The different colors correspond to the up and down temperature sweeps. The black trend is the plasmon energy calculated for crystalline Si thermal expansion.

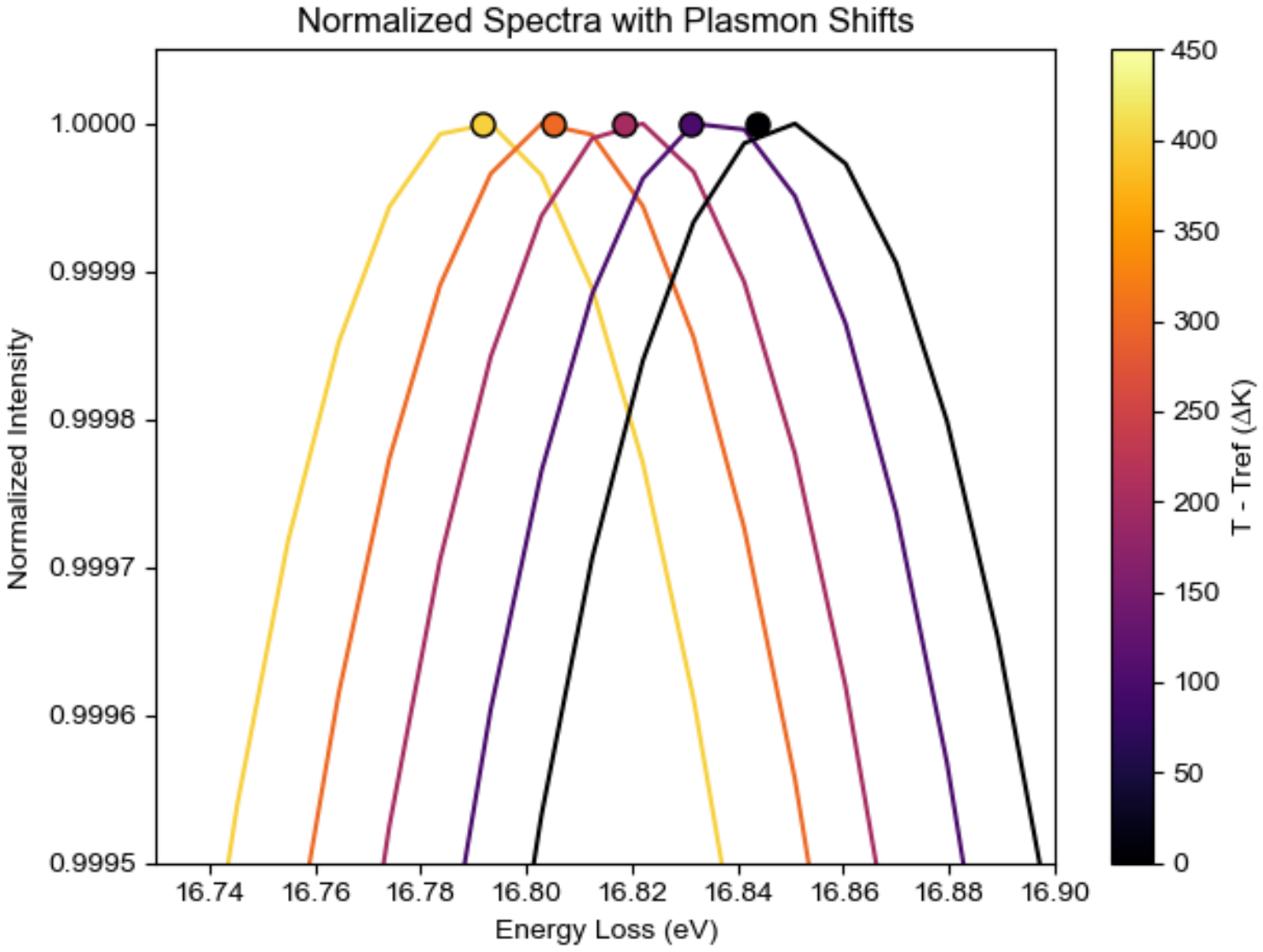

**Figure S6. Plasmon energies with Lorentzian-fit peak centers with temperature.** The temperature-dependent Lorentzian-fit plasmon energy in **Figure 1a**(inset) is compared to the average temperature-dependent plasmon spectrum in **Figure 1d**(inset). The plasmon energy (the scatter plot) is the average of the plasmon energies in **Figure S3**. The apparent fit deviation is likely due to data interpolation after fitting the spectra.

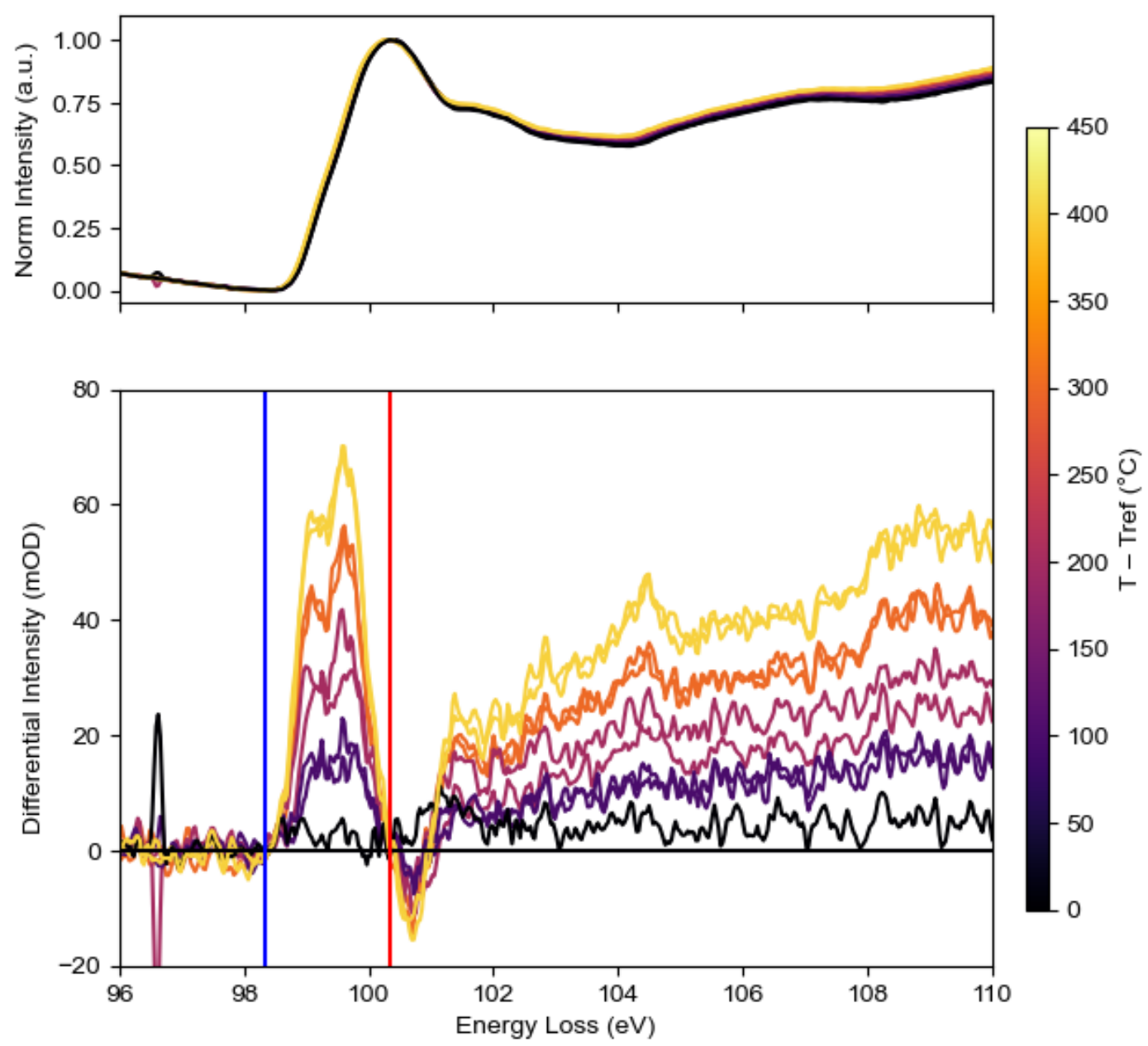

**Figure S7. Full Si $L_{2,3}$ differentials without smoothing.** To accurately visualize the differential spectra's signal-to-noise, the spectra are displayed without smoothing whereas smoothing is used for visualization in main text **Figure 2**. Each spectrum is the average spectrum across the STEM-EELS image at the temperature indicated by the color axis. The blue and red lines represent the minimum and maximum energies, respectively, used to normalize the ground-state spectra plotted above.

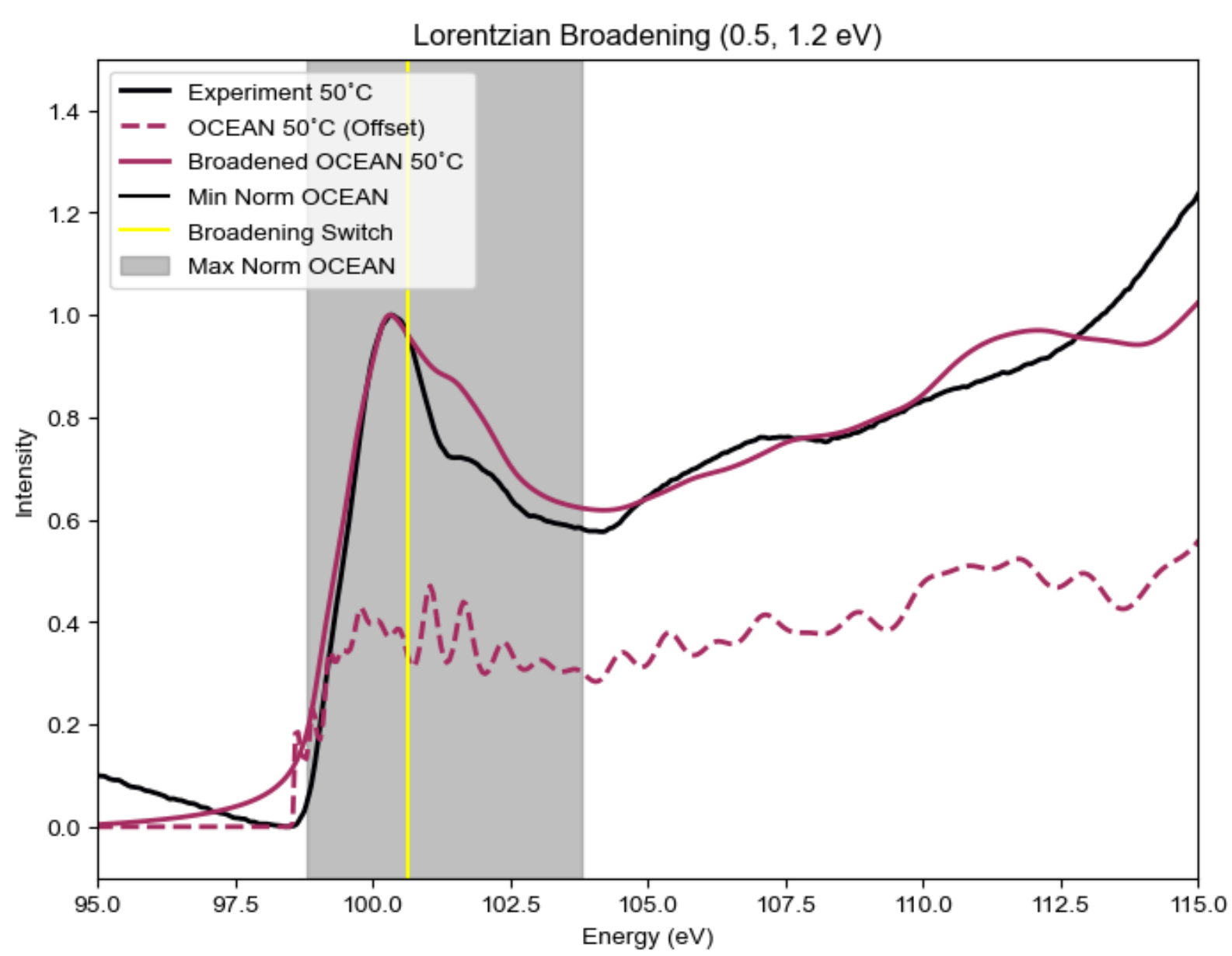

**Figure S8. Comparing calculated and experimental Si $L_{2,3}$ spectra.** To visualize the effects of energy-dependent Lorentzian broadening, the OCEAN output (before broadening) is displayed offset. The gray shaded region indicates the area used to normalize the OCEAN data (maximum intensity normalized to 1), and the yellow vertical line indicates the energy at which more Lorentzian broadening started to be applied (0.5 – 1.2 eV broadening applied linearly).

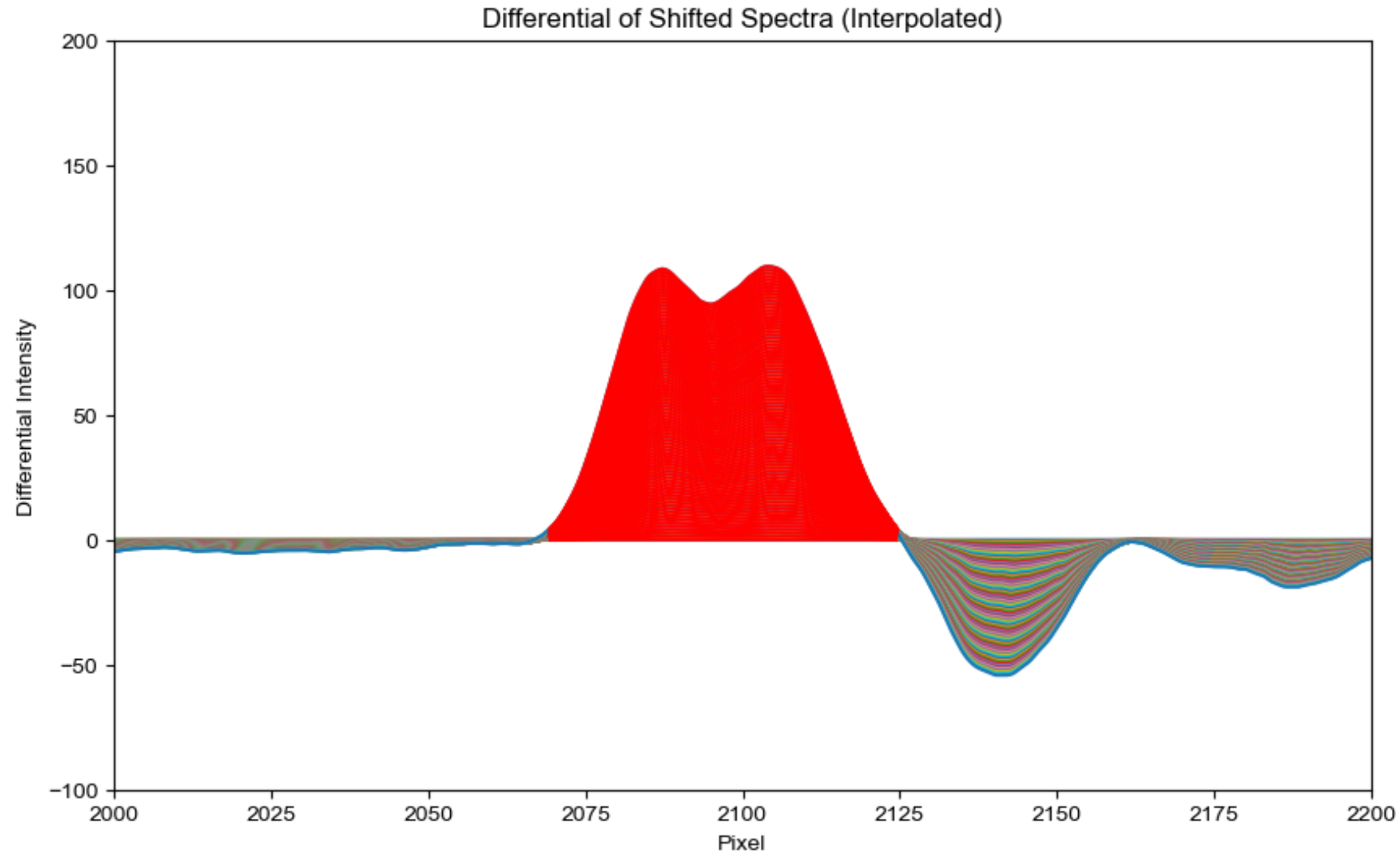

**Figure S9. Rigid spectral shift**. The rigid-shifted reference spectrum at $T_{ref}$ was shifted in 5 meV increments up to 105 meV. The maximum amplitude of the differential intensity (area analyzed fit in red) was then calculated and used to convert differential amplitude into meV spectral shift.

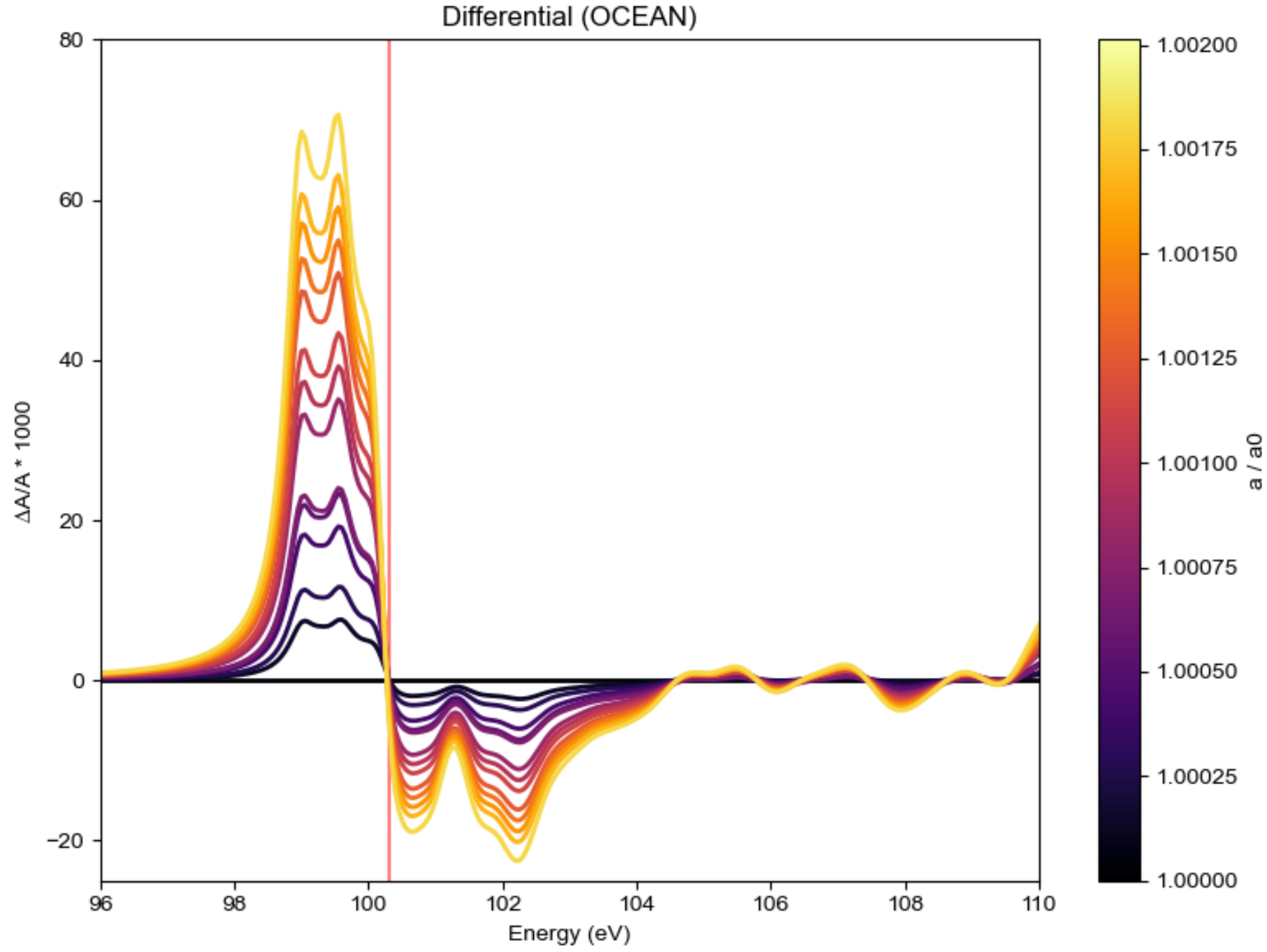

**Figure S10. Spectral normalization windows for Si $L_{2,3}$ calculated spectra.** The OCEAN-calculated Si L2,3 edge is plotted as a function of temperature. The red vertical line depicts the normalization window selected for the spectra after they are shifted as a function of temperature.

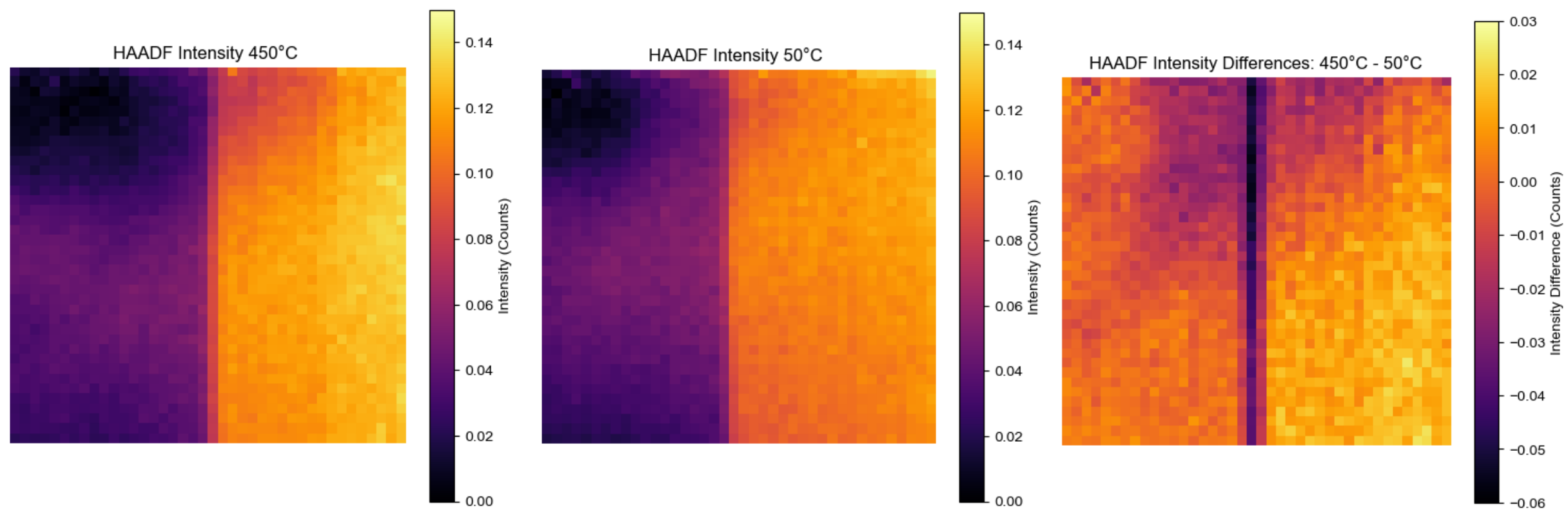

**Figure S11. HAADF intensity images.** The HAADF intensity at each pixel in the STEM images was also acquired and can be analyzed alongside the STEM-EELS data when looking for correlations in the datasets (see **Figure S14**). HAADF intensity is absolute, not normalized.

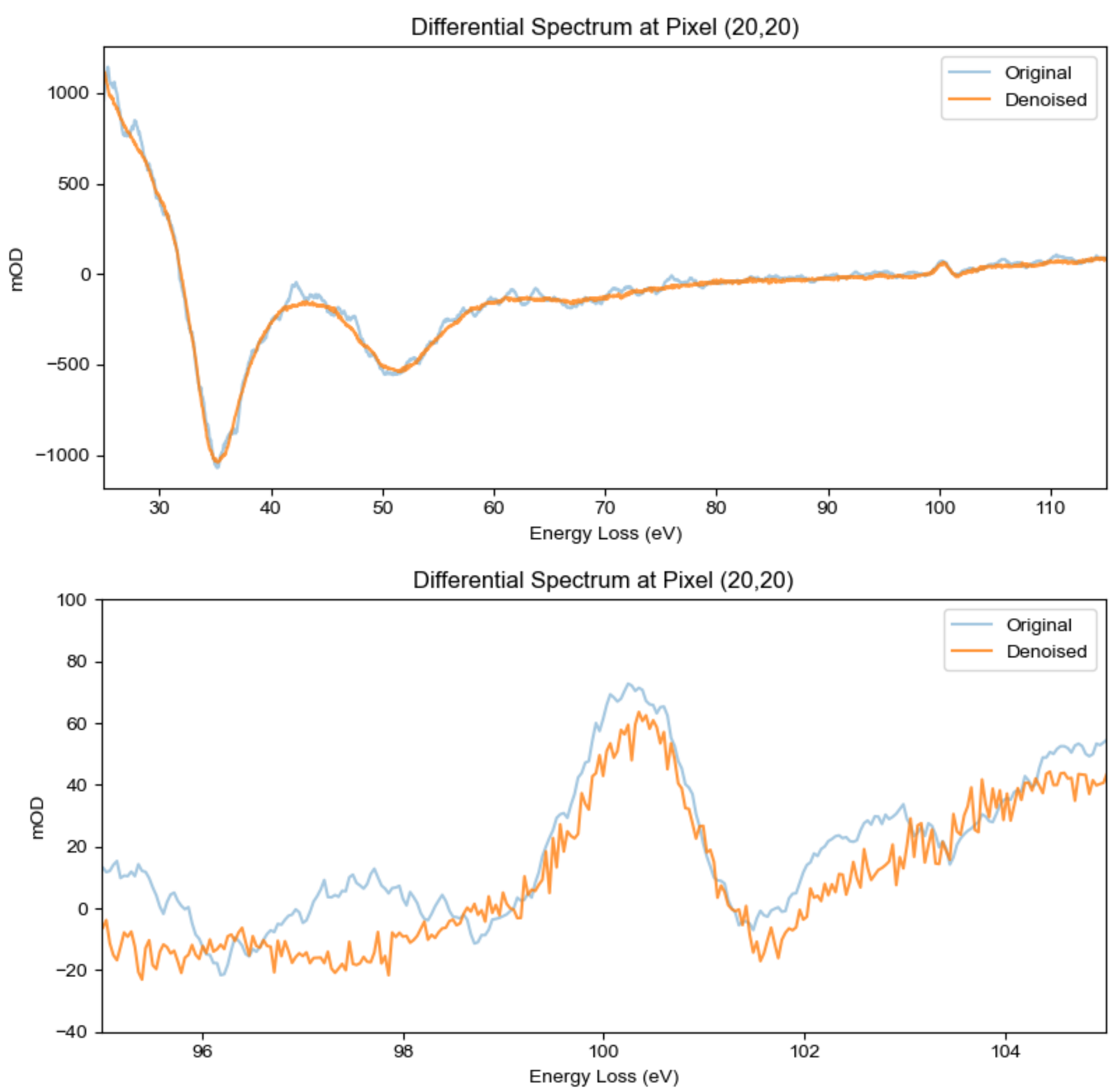

**Figure S12. Denoised example differential spectra.** The example spectrum of pixel 20,20 of the average 50˚C image is plotted. Denoising is performed after rigorous smoothing. Notably, the higher-frequency noise is amplified or reproduced while lower-frequency noise features are obscured. This results from the 3-step autoencoder used and the self-supervised learning of signal and noise from the test pixel set.

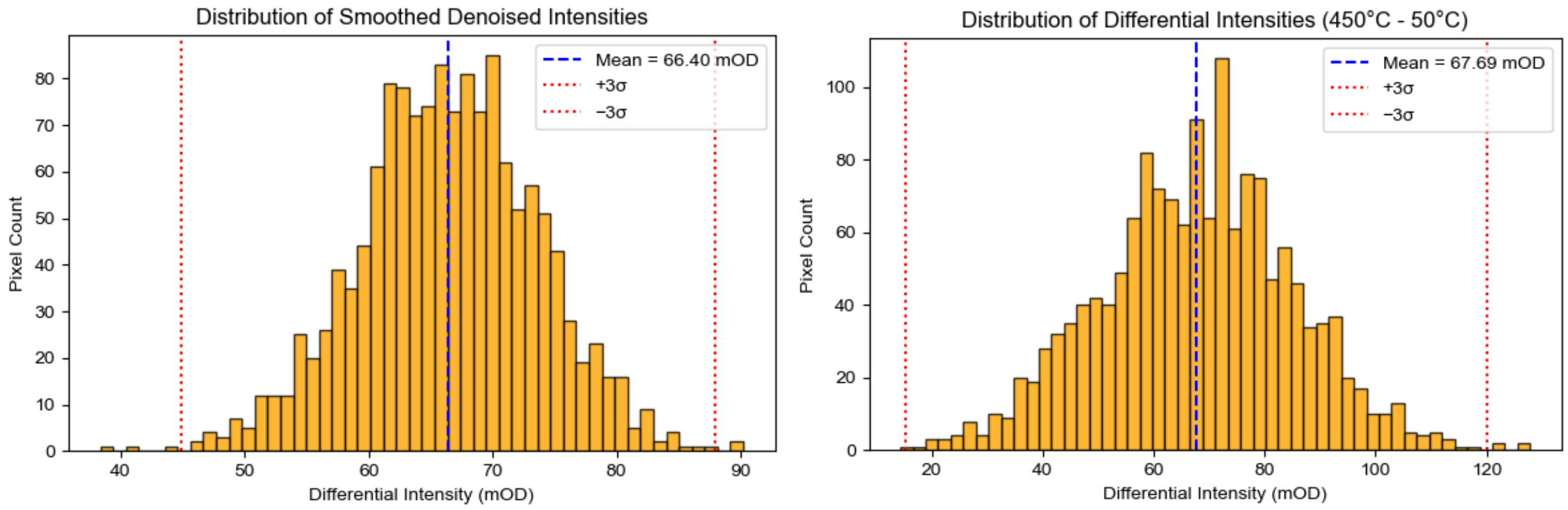

**Figure S13. Core-loss differential intensity statistics before (left) and after (right) denoising.** Each histogram represents the differential intensities from the core-loss thermometry images in the main text Figures 4c,d. The differential intensities are binned for each a.u. (mOD) differential intensity in the image. The mean and 3 standard deviations are plotted for visualization. Outliers are past the 3 standard deviation line and could be statistically removed if desired. See the exact statistics below:

**Denoising Statistics:**

Before Denoising:
Mean of differential image: 67.69 mOD
Standard deviation: 17.50 mOD
Percent deviation: 25.85%

After Denoising:
Mean of denoised image: 66.40 mOD
Standard deviation: 7.18 mOD
Percent deviation: 10.82%

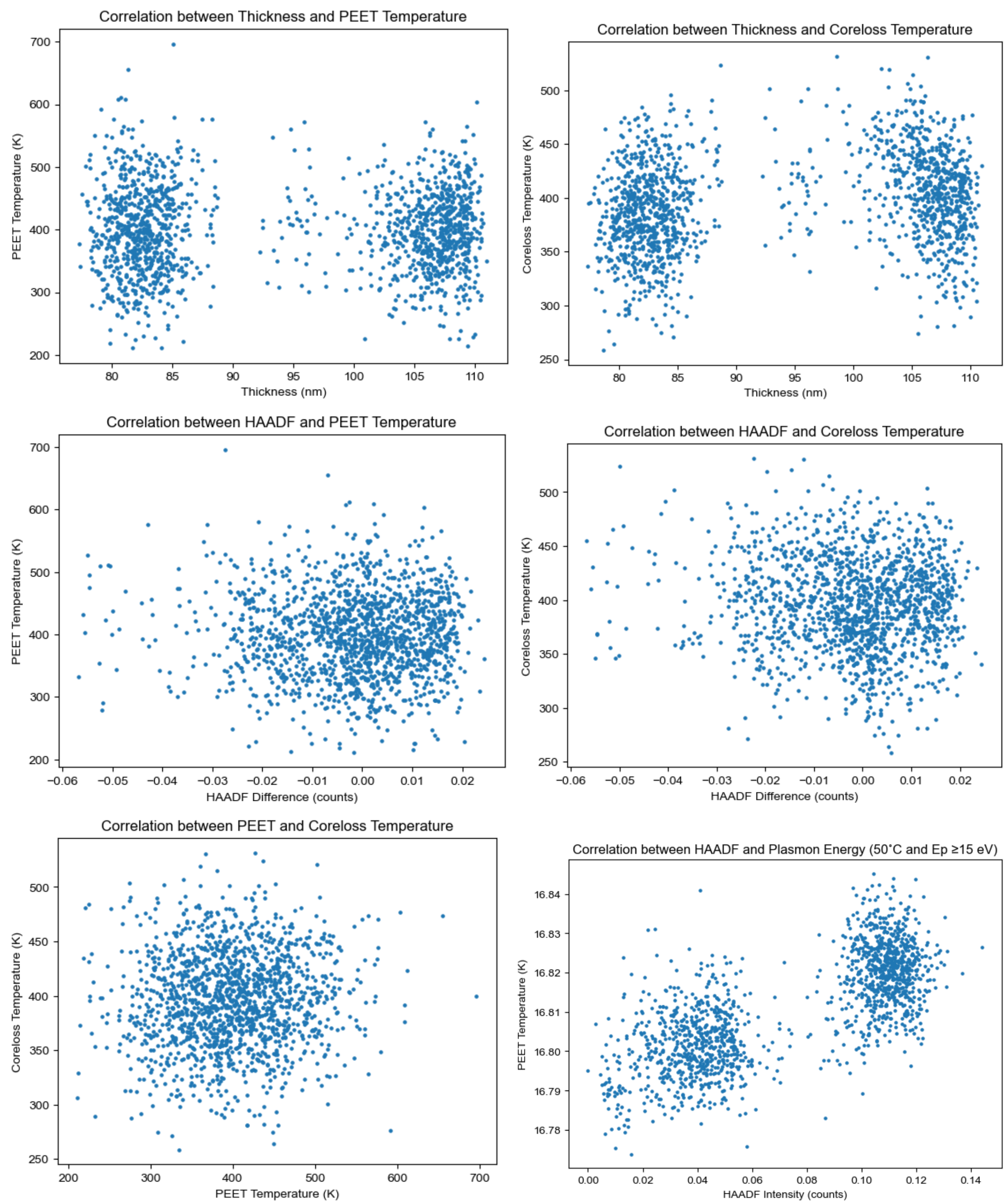

**Figure S14. Correlation plots.** Both Spearman and Pearson correlation algorithms were used to test for any correlations within the datasets. Correlation plots were made to better present these relationships. These correlations would potentially indicate pixel-by-pixel errors associated with the datasets and sources of uncertainty. A weak yet statistically significant correlation was only detected for the core-loss temperature and thickness, resulting from the variable multiple scattering background created by thickness dependence which influenced the denoising encoder.

[a]**<u>Correlation Results:</u>**
Pearson PEET and Thickness Correlation: -0.010 (p-value: 0.693)
Spearman PEET and Thickness Correlation: 0.014 (p-value: 0.574)

Pearson Coreloss and Thickness Correlation: 0.226 (p-value: 4.78e-19)
Spearman Coreloss and Thickness Correlation: 0.183 (p-value: 5.95e-13)

Pearson HAADF and PEET Correlation: -0.024 (p-value: 0.358)
Spearman  HAADF and PEET Correlation: 0.004 (p-value: 0.89)

Pearson HAADF and Plasmon Energy Correlation: 0.733 (p-value: 3.71e-256)
Spearman HAADF and Plasmon Energy Correlation: 0.688 (p-value: 5.4e-214)

[a]*The first number is the correlation, where +1 is a strong positive correlation. The second number is the p-value, where* $p<0.05$ *represents statistical significance.*

## STEM-EELS Data Analysis

**PEET Analysis Script**

```
## Calculate Plasmon Energy in each STEM-EELS Image Pixel, Generate PEET Imaging GIF

# === Constants ===
com1_window = 200 #ZLP fit window
com_threshold = 0
vmin_value = 16.75 #plotting color
vmax_value = 16.90
output_gif_path = os.path.join(directory, 'Figures', 'Plasmon_Lorentzian1_Gif.gif')

# === Lorentzian model ===
def lorentzian(x, amp, center, gamma):
    return amp * (gamma**2 / ((x - center)**2 + gamma**2))

def model_lorentzians(x, a1, c1, g1, a2, c2, g2, a3, c3, g3):
    return (
        lorentzian(x, a1, c1, g1) +
        lorentzian(x, a2, c2, g2) +
        lorentzian(x, a3, c3, g3)
    )

# === Sorting and prep ===
def extract_temp(file): return int(file.split('_')[0])
sorted_down = sorted(down_files, key=extract_temp, reverse=True)
sorted_up = sorted(up_files, key=extract_temp)
combined_files = sorted_down + sorted_up
combined_temp_C = [extract_temp(f) for f in combined_files]
temp_to_kref_map = {extract_temp(f): kref for f, kref in zip(down_files + up_files, temp_Kref)}
combined_temp_Kref = [temp_to_kref_map[t] for t in combined_temp_C]

# === Pixel-level processing function ===
def process_pixel(i, j, spectrum, pix_scale, common_energy_axis):
    # === Apply median filter
    spectrum = median_filter(spectrum, size=19)  # Outlier filter
    spectrum = savgol_filter(spectrum, window_length=17, polyorder=2)  # general smoothing

    energy = np.arange(len(spectrum)) * pix_scale
    peak_index = np.argmax(spectrum)

    com1_start = max(peak_index - com1_window // 2, 0)
    com1_end = min(peak_index + com1_window // 2, len(spectrum))
    zlp = spectrum[com1_start:com1_end]

    if np.sum(zlp) <= com_threshold:
        return i, j, np.nan, None
```

```
    com1 = np.sum(np.arange(com1_start, com1_end) * zlp) / np.sum(zlp)
    com1_eV = com1 * pix_scale
    shifted_energy = energy - com1_eV
    start_idx = com1_end + 200  # start 400 pixels after the ZLP COM window

    if start_idx >= len(spectrum):
        return i, j, np.nan, None  # avoid out-of-bounds

    fit_energy = shifted_energy[start_idx:]
    fit_spectrum = spectrum[start_idx:]

    if len(fit_energy) < 10 or np.max(fit_spectrum) < 1e-3:
        return i, j, np.nan, None

    p0 = [np.max(fit_spectrum) * 0.4, 16.8, 2.0,
          np.max(fit_spectrum) * 0.3, 33.6, 2.5,
          np.max(fit_spectrum) * 0.2, 50.4, 3.0]
    bounds_lower = [0, 16.8 - 1.5, 0.5, 0, 33.6 - 1.5, 0.5, 0, 50.4 - 1.5, 0.5]
    bounds_upper = [np.inf, 16.8 + 1.5, 6, np.inf, 33.6 + 1.5, 12, np.inf, 70, 100]

    try:
        popt, _ = curve_fit(model_lorentzians, fit_energy, fit_spectrum,
                            p0=p0, bounds=(bounds_lower, bounds_upper), maxfev=10000)
        interp_func = interp1d(shifted_energy, spectrum, kind='cubic', bounds_error=False,
fill_value=0)
        shifted = interp_func(common_energy_axis)
        return i, j, popt[1], shifted
    except RuntimeError:
        return i, j, np.nan, None

# === Process each temperature ===
plasmon_maps = []
avg_spectra_per_temp = []
spectra_label_map = []
frames = []

for temp_idx, (filename, temp) in enumerate(zip(combined_files, combined_temp_C)):
    file_path = os.path.join(directory, filename)
    dataset_list = hs.load(file_path)
    lowloss_signal = dataset_list[2]
    matrix_size = lowloss_signal.data.shape[:2]
    spectrum_length = lowloss_signal.data.shape[2]
    energy_axis = np.arange(spectrum_length) * pix_scale
```

```
    extra_range = 10  # You can adjust this depending on how much padding you want on either
side
    interp_factor = 4  # 4x more points for higher resolution
    num_interp_points = (spectrum_length + 2 * extra_range) * interp_factor

    common_energy_axis = np.linspace(
        energy_axis[0] - extra_range,
        energy_axis[-1] + extra_range,
        num=num_interp_points
    )

    plasmon_energy_map = np.full(matrix_size, np.nan)
    shifted_spectra = np.zeros((matrix_size[0], matrix_size[1], len(common_energy_axis)))

    pixel_tasks = [
        (i, j, lowloss_signal.inav[i, j].data.copy(), pix_scale, common_energy_axis)
        for i in range(2, matrix_size[0])  # skip top 2 rows
        for j in range(matrix_size[1])
    ]

    results = Parallel(n_jobs=-1)(
        delayed(process_pixel)(*task) for task in pixel_tasks
    )

    for i, j, center, shifted in results:
        plasmon_energy_map[i, j] = center
        if shifted is not None:
            shifted_spectra[i, j] = shifted

    plasmon_maps.append(plasmon_energy_map)
    avg_spectrum = np.mean(shifted_spectra[2:, :, :], axis=(0, 1))
    avg_spectra_per_temp.append(avg_spectrum)
    spectra_label_map.append((temp, avg_spectrum))

    # === GIF Frame ===
    fig, ax = plt.subplots(figsize=(6, 6))
    im = ax.imshow(plasmon_energy_map, cmap='inferno_r', vmin=vmin_value,
vmax=vmax_value)
    ax.axis('off')
    plt.colorbar(im, ax=ax, label='Plasmon Energy (eV)')
    fig.suptitle(f"T - Tref = {combined_temp_Kref[temp_idx]:.2f} K", fontsize=14, y=0.98)
    fig.tight_layout()
    canvas = FigureCanvas(fig)
    canvas.draw()
    image = np.frombuffer(canvas.buffer_rgba(), dtype='uint8')
    image = image.reshape(canvas.get_width_height()[::-1] + (4,))
```

```
    frames.append(image[:, :, :3])
    plt.close(fig)

# === Save outputs ===
plasmon_maps = np.array(plasmon_maps)
np.save(os.path.join(directory, 'plasmon_maps.npy'), plasmon_maps)
np.save(os.path.join(directory, 'plasmon_filenames.npy'), combined_files)

with open(os.path.join(directory, 'spectra_label_map.pkl'), 'wb') as f:
    pickle.dump(spectra_label_map, f)

if frames:
    imageio.mimsave(output_gif_path, frames, fps=2)
else:
    raise RuntimeError("No frames were collected.")
```

**Core-Loss Thermometry Analysis Script**

```
## Fit and align the STEM-EELS core-loss thermometry data

# Shift STEM-EELS spectra using only ZLP COM (no plasmon fitting)
com_ZLP_window = 200
extra_range = 5  # eV
avg_shifted_lowloss_per_temp = []
avg_shifted_coreloss_per_temp = []

for temp_idx, filename in enumerate(files_list):

    file_path = os.path.join(directory, filename)
    dataset_list = hs.load(file_path)

    lowloss_signal = dataset_list[2]
    coreloss_signal = dataset_list[3]

    matrix_size = lowloss_signal.data.shape[:2]
    energy_axis = np.arange(len(lowloss_signal.data[0, 0])) * pix_scale
    extended_energy_axis = np.linspace(energy_axis[0] - extra_range,
                                       energy_axis[-1] + extra_range,
                                       len(energy_axis) + 2 * extra_range)

    shifted_lowloss_spectra = np.zeros((matrix_size[0], matrix_size[1],
len(extended_energy_axis)))
    shifted_coreloss_spectra = np.zeros_like(shifted_lowloss_spectra)

    for i in range(matrix_size[0]):
        for j in range(matrix_size[1]):
            lowloss_data = lowloss_signal.inav[i, j].data
            coreloss_data = coreloss_signal.inav[i, j].data

            peak_idx = np.argmax(lowloss_data)
            zlp_start = max(peak_idx - com_ZLP_window // 2, 0)
            zlp_end = min(peak_idx + com_ZLP_window // 2, len(lowloss_data))
            x_zlp = np.arange(zlp_start, zlp_end)
            y_zlp = lowloss_data[zlp_start:zlp_end]

            if np.sum(y_zlp) > 0:
                com_ZLP = np.sum(x_zlp * y_zlp) / np.sum(y_zlp)
            else:
                com_ZLP = 0

            shift_eV = com_ZLP * pix_scale
            shifted_lowloss_spectra[i, j] = np.interp(extended_energy_axis,
```

```
                                                  energy_axis - shift_eV,
                                                  lowloss_data,
                                                  left=0, right=0)
            shifted_coreloss_spectra[i, j] = np.interp(extended_energy_axis,
                                                  energy_axis - shift_eV,
                                                  coreloss_data,
                                                  left=0, right=0)

    avg_shifted_lowloss = np.mean(shifted_lowloss_spectra, axis=(0, 1))
    avg_shifted_coreloss = np.mean(shifted_coreloss_spectra, axis=(0, 1))

    avg_shifted_lowloss_per_temp.append(avg_shifted_lowloss)
    avg_shifted_coreloss_per_temp.append(avg_shifted_coreloss)

print("DONE")
```

## Computational Methods

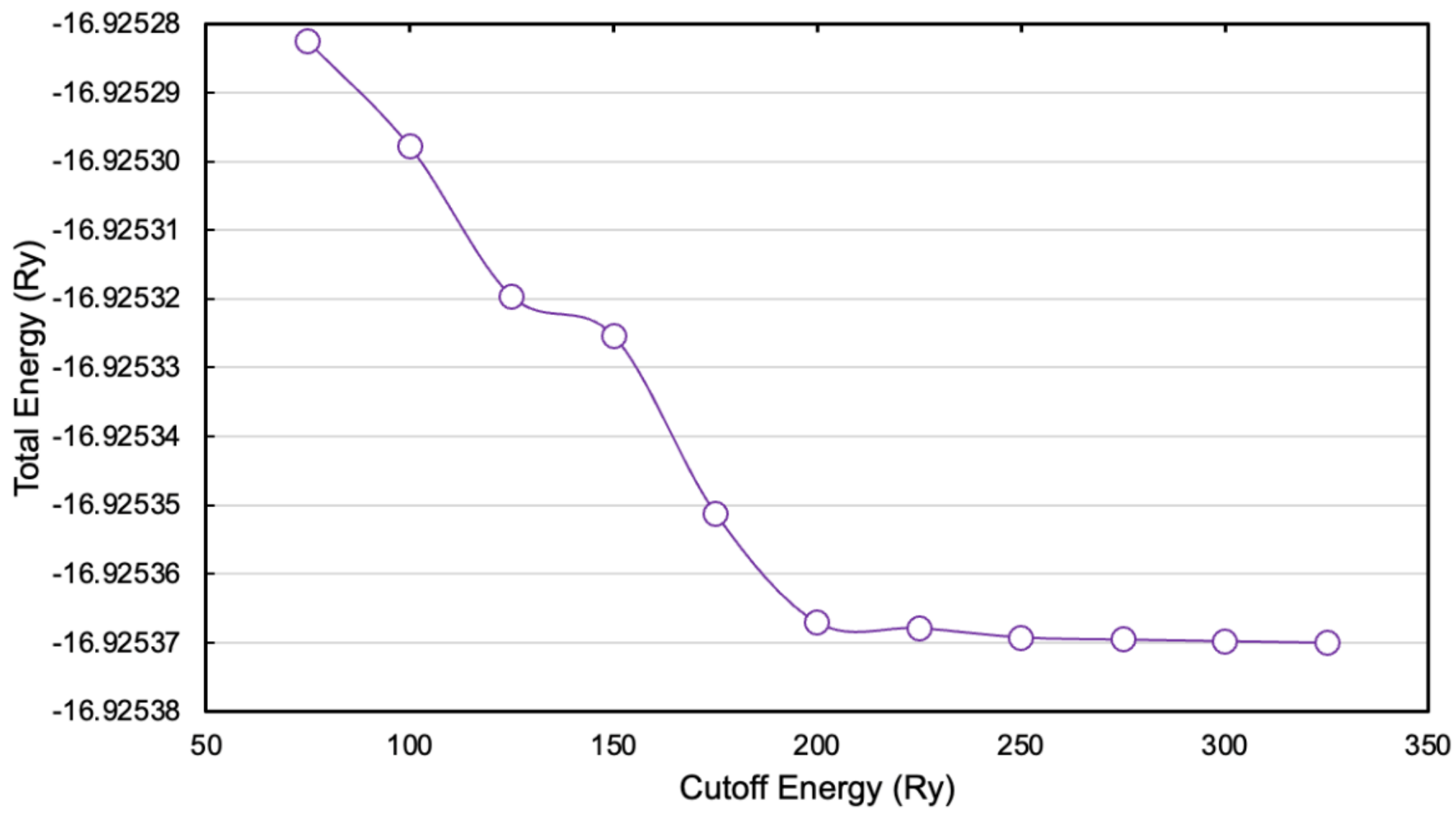

**Figure S14. Cutoff energy and energy optimization.** By iterating the self-consistent field calculation with varying wavefunction cutoff energies and at fixed convergence thresholds and *k*-mesh, the convergence of the calculation was found by minimizing the total energy. A 200 Ry cutoff energy was used for all subsequent calculations. See example input files.

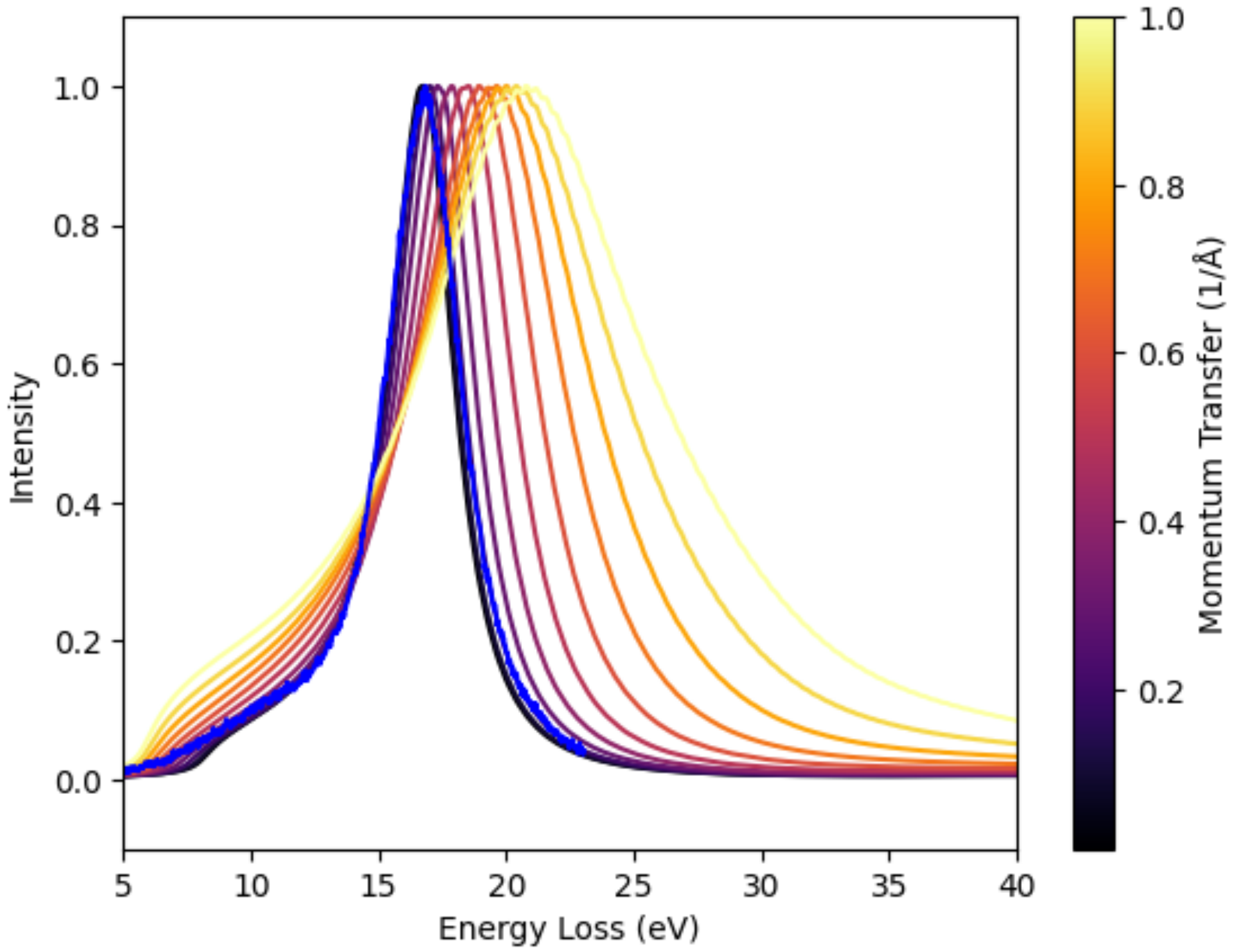

**Figure S15. turboEELS momentum dependence.** Momentum-dependent turboEELS spectra are compared to the experimental spectrum at the reference temperature (blue).

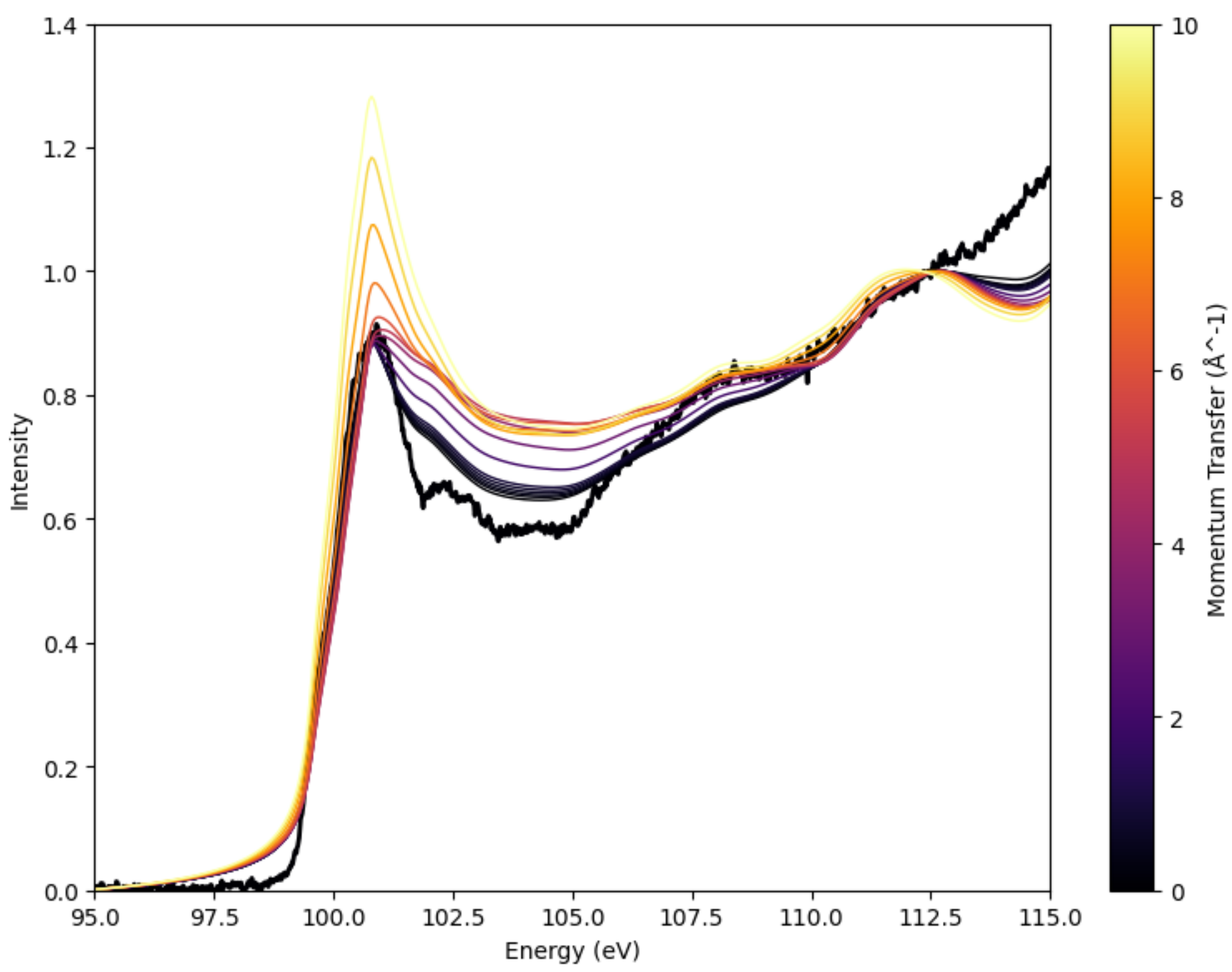

**Figure S16. Si $L_{2,3}$ OCEAN momentum dependence.** Momentum-dependent OCEAN spectra are compared to the experimental spectrum at the reference temperature (black).

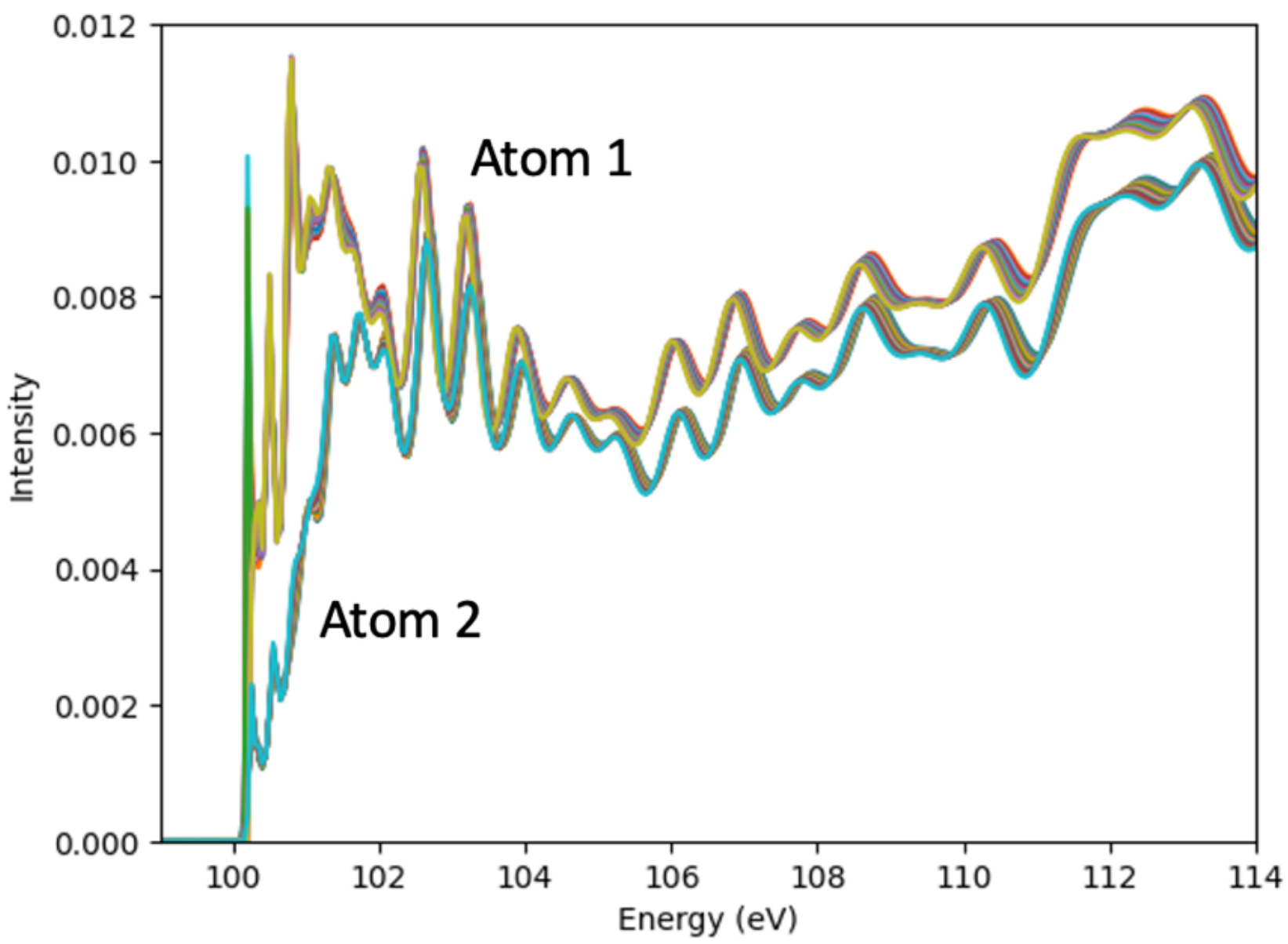

**Figure S17. Si $L_{2,3}$ OCEAN atom dependence.** There were two atoms in the Si unit cell. To make sure to avoid bias in the data analysis, both atoms were summed and averaged at each temperature.

**Scissor Shift Applied to OCEAN Data:**

**OCEAN SCF gap (eV)** [0.6748, 0.6755, 0.6761, 0.6768, 0.6775, 0.6782, 0.674, 0.6789, 0.6796, 0.6803, 0.681, 0.6817, 0.6823, 0.683, 0.6838]

**Predicted gap (eV)** [1.1400000000021464, 1.1317155479213157, 1.1245865472738859, 1.1179564608621686, 1.1115943432100084, 1.1053920357558942, 1.0992926300175108, 1.0932648755234582, 1.0872915832851213, 1.0813634448268268, 1.0754753647586335, 1.0696241704503942, 1.0638071937242035, 1.0580214632278906, 1.0522633477785055]

**OCEAN correction (eV):** [0.465 0.455 0.445 0.44 0.43 0.425 0.425 0.41 0.405 0.4 0.39 0.385 0.38 0.375 0.365]

**Simulated Lattice Parameters (OCEAN and turboEELS):**

| a/a0 | a (input Bohr) |
|---|---|
| 0.99986062 | 10.33 |
| 1 | 10.33144 = a0 |
| 1.00013938 | 10.33288 |
| 1.00027876 | 10.33432 |
| 1.00041814 | 10.33576 |
| 1.00055752 | 10.3372 |
| 1.0006969 | 10.33864 |
| 1.00083628 | 10.34008 |
| 1.00097566 | 10.34152 |
| 1.00111504 | 10.34296 |
| 1.00125442 | 10.3444 |
| 1.0013938 | 10.34584 |
| 1.00153318 | 10.34728 |
| 1.00167256 | 10.34872 |
| 1.00181194 | 10.35016 |

**Example turboEELS Input File:**

STEP 1) pw.x calculation (self-consistent field) ———————————————

```
&control
  calculation = 'scf'
  prefix = 'si'
  pseudo_dir = './'
  outdir = './out'
  wfcdir = 'undefined'
  tstress = .false.
  tprnfor = true
  wf_collect = .true.
  disk_io = 'low'
/
&system
  ibrav = 2
  celldm(1) = 10.33 ! loops lattice param here!
  nat = 2
  ntyp = 1
  noncolin = .false.
  lspinorb = .false.
  ecutwfc = 200
  occupations = 'fixed'
  smearing = 'gaussian'
  degauss = 0.02
  nspin  = 1
  tot_charge = 0
  nosym = false
  noinv = false
  nbnd = 6
/
&electrons
  conv_thr = 1e-10
  mixing_beta = 0.7
  electron_maxstep = 100
  startingwfc = 'atomic+random'
  startingpot = 'atomic'
  diagonalization = 'david'
/
ATOMIC_SPECIES
  Si  28.0855  Si.UPF
ATOMIC_POSITIONS {alat}
  Si  0.00    0.00    0.00
  Si  0.25    0.25    0.25
K_POINTS {automatic}
16 16 16 1 1 1
```

STEP 2) turbo_eels.x calculation (linear-response TDDFT) ———————————————

```
&lr_input
   prefix='si',
outdir='./out',
   restart_step = 250,
   restart = .false.
/
&lr_control
   itermax = 500,
   q1 = 0.200,
   q2 = 0.000,
   q3 = 0.000,
/
```

STEP 3) turbo_spectrum.x calculation (post-processing the spectrum) ————————————

```
&lr_input
   prefix='si',
   outdir='./out',
   eels = .true.
   itermax0 = 500
   itermax  = 20000
   extrapolation = "osc"
   epsil = 0.03
   units = 1
   start = 0.0d0
   increment = 0.01d0
   end = 50.0d0
   verbosity = 0
/
```

**Example OCEAN Input File:**

```
################################
# SETUP
dft.program { qe }
computer.para_prefix{ mpirun -n 32 }

# ppdir { '../' }

################################
# DFT AND SCREENING
dft.ecut 200
dft.toldfe 1d-10
dft.tolwfr 1d-10
dft.den.kmesh{ 16 16 16 } # ngkpt mesh for ground-state calc
dft.nstep 100
dft.mixing 0.7

screen.kmesh{ 4 4 4 }
screen.shells{ 3.5 } # radius for shell for screening
screen.nbands 100

################################
# ATOMIC STRUCTURE
structure.znucl{ 14 }
structure.typat{ 1 1 }
structure.epsilon 11.4  # Static dielectric const

acell{ 10.33 10.33 10.33 }
structure.rprim{
0.5 0.5 0.0
0.5 0.0 0.5
0.0 0.5 0.5 }
structure.xred{
0.00 0.00 0.00
0.25 0.25 0.25 }

################################
# BSE AND XAS CALCULATION
bse.nbands 100
bse.xmesh{ 6 6 6 }
bse.kmesh{ 12 12 12 } # nkpt for final state
bse.core.broaden{ 0.000001 } # Lorentzian broadening (eV)

calc.mode xas
calc.edges{ -14 2 1 } # Si L23
```